\documentclass[sigconf,screen,urlbreakonhyphens]{acmart}

\AtBeginDocument{%
  }

\usepackage{tikz}
\usepackage{amsmath}
\usepackage{graphicx}
\graphicspath{ {./images/} }
\usepackage{longtable}
\usepackage{csquotes}
\usepackage{appendix}
\usepackage[utf8]{inputenc}
\usepackage{xspace} 
\usepackage{hyperref}
\usepackage{xcolor}
\usepackage{colortbl}
\usepackage[utf8]{inputenc}
\definecolor{color1}{RGB}{63,87,101}
\definecolor{color2}{RGB}{213,232,212}

\newcommand{\KidsList}{\textsc{Children YTB Kids List}\xspace}

\newcommand{\SeedList}{\textsc{Children Seed List}\xspace}
\newcommand{\ControlList}{\textsc{Adult List}\xspace}

\setcopyright{acmcopyright}
\copyrightyear{2018}
\acmYear{2018}
\acmDOI{XXXXXXX.XXXXXXX}

\acmConference[Conference acronym 'XX]{Make sure to enter the correct
  conference title from your rights confirmation emai}{June 03--05,
  2018}{Woodstock, NY}
\acmPrice{15.00}
\acmISBN{978-1-4503-XXXX-X/18/06}

\copyrightyear{2023} 
\acmYear{2023} 
\setcopyright{acmlicensed}\acmConference[CCS '23]{Proceedings of the 2023 ACM SIGSAC Conference on Computer and Communications Security}{November 26--30, 2023}{Copenhagen, Denmark}
\acmBooktitle{Proceedings of the 2023 ACM SIGSAC Conference on Computer and Communications Security (CCS '23), November 26--30, 2023, Copenhagen, Denmark}
\acmPrice{15.00}
\acmDOI{10.1145/3576915.3623172}
\acmISBN{979-8-4007-0050-7/23/11}

\begin{document}

%\title{Are children targeted with ads? \\ An investigation on YouTube}
\title{Are children targeted with ads? \\ An investigation on YouTube}
\title{Can advertisers target their ads to children? \\ An investigation on YouTube}
%\title{Contextual advertising as a way to target children with ads\\ An investigation on YouTube}
%\title{Contextual advertising enables targeting children with ads}
%\title{Investigating targeting mechanisms to target children with ads}
%\title{Targeting children with ads: mechanisms and legal considerations}
\title{A legal and technical analysis of targeting children with ads}
\title{Advertisers can target their ads to children: a technical and legal investigation}
\title{How online advertising enables marketing to children}
\title{Marketing to Children Through Online Targeted Advertising: Targeting Mechanisms and Legal Aspects}

%\title{How advertisers can target their ads to children}
\author{Tinhinane Medjkoune}
\affiliation{
\institution{Univ. Grenoble Alpes, CNRS, Grenoble INP, LIG} \country{France}}
%\email{tinhinane.medjkoune@univ-grenoble-alpes.fr}
\author{Oana Goga}
%\email{oana.goga@cnrs.fr}
\affiliation{%
  \institution{LIX, CNRS, Inria, Ecole Polytechnique, Institut Polytechnique de Paris}
  \country{France}}
\author{Juliette Senechal}
\affiliation{
\institution{Universit\'e de Lille, CRDP, DReDIS-IRJS}
\country{France}}

\renewcommand{\shortauthors}{Tinhinane Medjkoune, Oana Goga, \& Juliette Senechal}
%% use of ampersand (\&) versus ''and'' is to saves space.

% d\renewcommand{\shortauthors}{Trovato et al.}

\begin{abstract}
Many researchers and organizations, such as WHO and UNICEF, have raised awareness of the dangers of advertisements targeted at children.
While most existing laws only regulate ads on television that may reach children, lawmakers have been working on extending regulations to online advertising and, for example,  forbid (e.g., the DSA) or restrict (e.g., the COPPA) advertising based on profiling to children.

At first sight, ad platforms such as Google seem to protect children by not allowing advertisers to target their ads to users that are less than 18 years old. However, this paper shows that other targeting features can be exploited to reach children. For example, on YouTube, advertisers can target their ads to users watching a particular video through placement-based targeting, a form of contextual targeting. Hence, advertisers can target children by simply placing their ads in children-focused videos. Through a series of ad experiments, we show that placement-based targeting is possible on children-focused videos and, hence, enables marketing to children. 
In addition, our ad experiments show that advertisers can use targeting based on profiling (e.g., interest, location, behavior) in combination with placement-based advertising on children-focused videos.  
We discuss the lawfulness of these two practices with respect to DSA and COPPA. 

Finally, we investigate to which extent real-world advertisers are employing placement-based targeting to reach children with ads on YouTube. We propose a measurement methodology consisting of building a Chrome extension able to capture ads and instrumenting six browser profiles to watch children-focused videos. Our results show that 7\% of ads that appear in the children-focused videos we test use placement-based targeting. Hence, targeting children with ads on YouTube is not only hypothetically possible but also occurs in practice.  
We believe that the current legal and technical solutions are not enough to protect children from harm due to online advertising. A straightforward solution would be to forbid placement-based advertising on children-focused content.

\end{abstract}




% %%
% %% Keywords. The author(s) should pick words that accurately describe
% %% the work being presented. Separate the keywords with commas.
% %% A "teaser" image appears between the author and affiliation
% %% information and the body of the document, and typically spans the


\maketitle

%!TEX root = main.tex

\section{Introduction}

%\oana{clar our terminology target: I want to reach children and I can; contextual -- behaviora/profiling -- based on history}
Many researchers have studied how advertisements affect children since their television appearance and highlight their potential negative impact on children's development and well-being~\cite{ijrsi18}. 
One reason why advertisements can harm children is that their cognitive skills are underdeveloped. Children cannot distinguish an advertisement from a television program before age three and hence cannot fire the mental defenses adults have when seeing ads~\cite{pretvad}. Only between the ages of 8 and 12 do children begin to understand the commercial ambitions of ads ~\cite{adsRecognition}. Nevertheless, even if they understand the underlying commercial intent, the attractive nature of advertisements makes children infatuated with them (see Section~\ref{sec:rw}). As WHO \& UNICEF put it in their 2020 report~\cite{unicefwho20}: ``{\em Marketing to children is deliberate, strategic, innovative and well resourced, and exploits their developmental vulnerability.''}

Understandably, many laws across the World forbid or restrict children's advertising. For example, in France, Article 7 of the decree of March 27, 1992, states that ``{\em advertising must not cause moral or physical harm to minors [...] it must not directly encourage minors to purchase a product or service by exploiting their inexperience or credulity [...]}'' and Article 15 of the same decree prohibits advertising breaks on television during children's programs of less than 30 minutes~\cite{decretFR}. While most existing laws have been set to protect children from television ads, legislators worldwide have been working on updating legislation to consider ads on social media and streaming platforms.

% However, most existing laws have been set to protect children from ads on television, and they do not apply to ads on social media and streaming platforms.  
 %Sweden goes even further, and simply forbids the broadcasting of any television commercials aimed at children under 12~\cite{swedenlaw}. 

Given the gravity of the problem, in this paper, we focus on \emph{understanding whether children can be targeted with ads on online streaming platforms}.\footnote{By \emph{targeting children}, we refer to any means an advertiser can exploit features provided by online advertising platforms to reach an audience made mostly of children; hence,  targeting does not necessarily mean micro-targeting or profiling.}
For clarity, our work focuses not on whether children accidentally see ads while watching videos on online streaming platforms but on whether advertisers can intentionally target them with ads. 
We take YouTube as a case study as it is one of the most used online platforms for children~\cite{statista, pew}.  
Understanding how and if advertisers can reach children is essential for understanding what risks targeting technologies bring to children, understanding what transparency we need, and helping inform lawmakers with adapted restrictions. 

%Regulators across the World are working on updating legislation to consider the case of online advertising. 
We review in Section~\ref{sec:bkg_law} legal acts that relate to advertising to children in the U.S. and Europe to understand what is currently legally allowed, forbidden, or restricted. In particular, we examine the Children's Online Privacy Protection Rule (COPPA) in the U.S.~\cite{coppa} and the Digital Services Act, a new Regulation voted on 19 October 2022 at the European Union level~\cite{DSA}. The COPPA Act does not prohibit advertising to children; however, it places some restrictions on it: it states that advertisers and content owners may not collect any personal information (which includes cookies and other persistent identifiers) from children under 13 years of age without verifiable parental consent. Hence, COPPA restricts online platforms' capabilities to serve profile-based ads to children but does not restrict contextual-based advertising.  
The DSA takes a step further and forbids altogether targeting children with ads based on profiling (e.g., interest, location, behavior).  % We present an in-depth analysis of these legal acts in Section~\ref{sec:bkg_law}. 

%\update{For example, the COPPA Rule in the US allows targeting of children through contextual advertising and states that advertisers and content owners may not collect any personal information from children under 13 years of age without verifiable parental consent. This is intended to stop the behavioral advertising, retargeting, and profiling of children under 13. Moreover, a new Regulation, at the European Union level, of 19 October 2022, the Digital Services Act, has made a further step forward and forbids directly ads based on profiling at children.} 

%With more than 2.2 billion users, YouTube is among the most used streaming platforms. While ads on online streaming platforms are similar in format with ads on television, there are almost no laws that regulate advertising to children on online streaming platforms, with most existing laws only regulating advertising on television.  

%is regulating how children's personal information is processed to protect them from targeted advertising, and the European Commission is working on a set of laws to regulate ads targeted at children 

On the technical side, ad platforms have also deployed solutions to protect children from online marketing. For example, the advertising interface on Google does not allow advertisers to target users less than 18 years old~\cite{GoogleAdsAge}. 
In addition, Google launched YouTube Kids in 2015. This platform only contains content (videos and ads) filtered explicitly for children. Content creators and advertisers cannot explicitly ask for their videos or ads to be shown on YouTube Kids. YouTube curates content in an independent way (see Section~\ref{sec:background})~\cite{ykads}. 
As there is no direct way to target children on YouTube Kids, this paper focuses on targeting mechanisms that allow advertisers to target children on YouTube. This choice is consequential because content available on YouTube Kids is also available on YouTube, and previous studies showed parents use YouTube more frequently than YouTube Kids to let their children watch videos~\cite{YKDef}.

%Nevertheless, while no direct way in which advertisers can target children on YouTube Kids, the content available on YouTube Kids is also available on YouTube; and hence, advertiser 
%Therefor, in this work we focus on YouTube 
%
%Hence, there is hile YouTube Kids is widely adopted, still more parents say they are using YouTube to let
%

%However, we show in this paper that other targeting features can be exploited to reach children.

First, we investigate what targeting mechanisms can be exploited to target children with ads on YouTube (see Section~\ref{sec:targeting}). 
We find that, Google allows advertisers to target their ads to users watching a \emph{particular} video through \emph{placement-based targeting}, a form of contextual advertising. Hence, advertisers could target children by simply placing their ads in children-focused videos.  
Through a series of real-world ad experiments, we show that placement-based targeting is allowed on a set of children-focused videos we curated from YouTube Kids, hence, enabling marketing to children. In addition, through four other ad campaigns, we show that placement-based targeting on children-focused videos can be used in combination with targeting based on profiling.\footnote{For example, target users interested in Sports that are watching the video ``Peppa Learns How To Whistle.''} %Hence, it is possible to target children with ads online.  
According to our legal analysis, placement-based targeting on children-focused videos is not forbidden by either the COPPA or the DSA; however, the use of profiling for targeting children is forbidden in the DSA and might be (as the text is not clear in this respect) forbidden in the COPPA as well. Hence, combining placement-based targeting with targeting based on profiling is against these rules. 

%\update{Our experiments show that there is a need to better distinguish between cases where targeting is carried out solely on the basis of context or where targeting is carried out by mixing contextual placement and profiling. This distinction is necessary to effectively implement the European (DSA) and American (COPPA) texts protecting children against advertising based on profiling.}

%Hence, combining placement-based targeting on children-focused videos with profile-based targeting might be interpreted as problematic in the two acts. 

%his distinction is necessary to effectively implement the European (DSA) and American (COPPA)

%To validate whether in practice such targeting is allowed on children-focused videos, we created three ad campaigns that target a set of children-focused videos curated from YouTube Kids. All our ad campaigns campaigns were validated by Google and reached users. In addition,  through four other ad campaigns we show that placement-based targeting can be used in combination with targeting based on profiling.
%Hence, current protection mechanisms provided by ad platforms can simply be circumvented to reach children. In addition, the use of profiling for targeting children is forbidden in the DSA and is contrary to what Google publicly claimed.

Second, we investigate if advertisers exploit these targeting mechanisms to reach children in the wild (see Section~\ref{sec:metho}). For this, we propose a measurement methodology that consists of building a Chrome extension that can collect data about the ads shown on YouTube videos and instrument six browser profiles to watch children-focused videos. Using this approach, we collected 3,321 ads that appear on 620 videos. Next, given an ad shown on a particular video, the challenge is to distinguish whether the ad was targeted based on contextual parameters (e.g., placement-based), based on profiling parameters (e.g., behavior, interest, location), or based on other parameters such as time and language. 
We find a simple yet effective way to make the distinction by exploiting the ad explanations provided in the ``Why you're seeing this ad'' feature~\cite{myadcenter}. 
Our analysis shows that 7\% of the ads we collected while watching two sets of children-focused videos (across the six profiles) had an ad explanation suggesting the advertiser used placement-based targeting. In addition,  25\% of ads collected on children-focused videos had ad explanations suggesting they have been targeted based on profiling. 
While these numbers are not representative and are obtained over a small scale dataset, they show that targeting children is not only hypothetically possible but also appears to occur in practice.

%\oana{maybe present results per list to show that the percentage is higher for children-focused videos that are not sourced from YouTube Kids}
Finally, we observe that YouTube provides a reduced level of transparency for ads on YouTube videos labeled as being suitable for children\footnote{These YouTube videos have a redirection to YouTube Kids as shown in Figure~\ref{fig:youtube-kids-redirection} in appendix just below the video content.} compared to other videos (see Section~\ref{sec:res}).  
To validate this, we created an ad campaign and instructed Google to use placement-based targeting and show it on ten videos sourced from YouTube Kids and ten adult-focused videos. Whenever our ad was shown on adult-focused videos, it had the ad explanation, ``The video you are watching.'' that reflected precisely our targeting parameters. However, on the children-focused videos sourced from YouTube Kids, the explanation was: ``Personalization is disabled for this account or content. Therefore, this ad is not personalized based on your data. Its distribution depends on other factors (such as the time or your geographical position)''. While this explanation is not necessarily wrong, it does not reflect the precise targeting parameters we used in our ad campaign. 
Hence, for the same ad campaign (and the same targeting parameters) and the same profile watching the video, Google showed different ad explanations depending on whether the video was present on YouTube Kids or not. 
Therefore, not only can children be targeted with ads through placement-based advertising, the ad explanations provided on children-focused videos do not reflect the use of such targeting.

The implications of this study are multi-faceted. Since many countries are updating their legislation to protect children from online harm,  we hope our paper clarifies how advertisers can reach children online and provides valuable technical insights. In particular, we want to raise awareness that advertisers can reach children through placement-based advertising. While the DSA and the COPPA do not currently forbid this practice, we think regulators have not realized that contextual advertising can be done at the granularity of the video the user is watching and, hence, can be used to reach children watching well chosen sets of videos that could be indicators of their age, gender, or interests. %In addition, current online legislation 
Even worse, current online advertising legislation is missing clear guidelines on the ad content that can be targeted at children (as in the case of television advertising); this is important to clarify, especially since placement-based advertising is currently allowed. 

We believe that the current legal and technical solutions are not enough to protect children from harm due to online advertising. We \emph{recommend} that: (1) Regulators and ad platforms should simply forbid placement-based advertising on children-focused content. (2) Regulators should set clear guidelines on the ad content that can be shown on children-focused content, including ads that are shown because they were targeted at children-focused content or ads that appear there by chance and use other targeting parameters.  (3) When addressing Article 39 (Additional online advertising transparency) in the DSA, ad platforms should include, for all ads, whether the advertiser used placement-based targeting and provide the complete list of placements used by the advertiser. This will enable researchers to check if the ads' content is harmful to the chosen audience. (4) Ad platforms should provide consistent ad explanations for the same ad. Ad explanations should not be generated on the fly based on the content watched or the profile of the user watching. In addition, we believe that the current ad explanations provided by Google on children-focused videos are misleading and not ``meaningful''  as demanded in Article 26 of the DSA. 

At last, we want to salute efforts from online platforms to provide data about how their systems work, including explanations of why users receive particular ads and detailed statistics on ad delivery. Without such platform-provided data, this auditing study could not have been performed. 

\section{Background}
\label{sec:background}

\subsection{Advertising on YouTube}
Google provides seven broad categories of targeting for advertising on YouTube~\cite{GoogleAdsTargeting}:

\vspace{0.5mm}
\noindent $\bullet$ \textbf{Location-based:} 
Advertisers can specify the country, city, postal codes, or a radius as small as 1 km around an address. 

\vspace{0.5mm}
\noindent $\bullet$ \textbf{Demographic-based:}  Advertisers can select over 440 combinations of demographic groups based on gender, age, parental status, and household income. 

\vspace{0.5mm}
\noindent $\bullet$ \textbf{Behavioral or interest-based:} Advertisers can select people with particular interests (such as technology), habits (such as Cinema enthusiasts), or Market Audience and Life Events. 

\vspace{0.5mm}
\noindent $\bullet$ \textbf{Re-targeting:} Re-targeting allows advertisers to reach on YouTube, the users that visited their websites or application (outside Google or YouTube). Advertisers can also upload a CSV file that contains the contact information (e.g., emails, phone numbers, physical addresses) of the users they want to reach. This data is typically collected through loyalty cards. 

\vspace{0.5mm}
\noindent $\bullet$ \textbf{Keyword-based:} Advertisers can specify a list of keywords, and their ads will appear when users search for the particular keywords on YouTube.   

\vspace{0.5mm}
\noindent $\bullet$ \textbf{Theme-based:} This allows advertisers to place their ads on videos related to a specific topic. There are over 26 proposed topics ranging from real estate to art. 

\begin{figure}[t]
        \centering
        \includegraphics[scale=0.33]{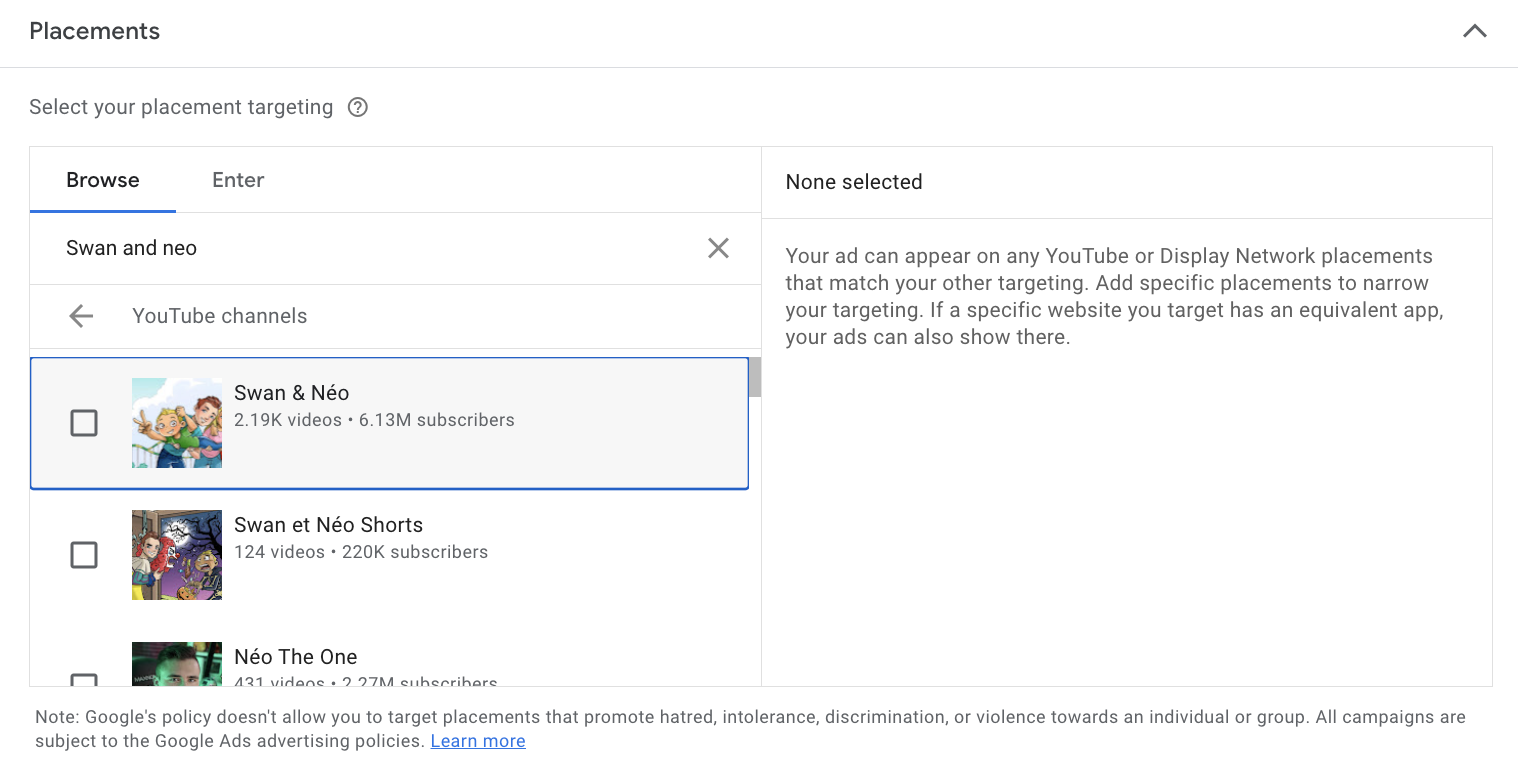}
          %      \vspace{-3mm}
        \caption{Placement-based targeting in Google.}
        \label{fig:Contextual-based}
        \vspace{-5mm}
    \end{figure}

\vspace{0.5mm}
\noindent $\bullet$ \textbf{Placement-based:} This targeting category is the most interesting to our study. Google allows advertisers to specify precisely on which YouTube channels or videos they wish their ads to appear on (see  Figure~\ref{fig:Contextual-based}). 
For example, an advertiser can specify their ad should only appear on the "Food Court: France Edition" video. 

Both \emph{theme-based advertising} and \emph{placement-based advertising} are two forms of \textbf{contextual-based advertising}, as the targeting is based only on information related to the content and not the user. For placement-based targeting, the advertiser specifies precisely the placement of the ad; hence, the ad will then appear only on the specific videos selected by the advertiser. For theme-based targeting, the advertiser just specifies the theme, and the ad platform decides what videos are related to the theme selected by the advertiser. 

\emph{Demographic, behavioral, interest-based} as well as \emph{re-targeting} are forms of \textbf{profiling-based advertising}, as the targeting is based on information gathered on the (historical) activity of the user. This targeting is also sometimes called \textbf{personalized advertising}. 

These targeting categories can be combined when placing an ad. In addition, advertisers can set the language, the time when to broadcast the ad and the devices used by the users they want to reach. Google imposes a minimum reach requirement, and ads will not be delivered if Google estimates they have an audience of less than 1000 users. 
Finally, Google proposes an ``audience expansion'' feature that gives Google the power to expand the targeting to additional relevant audiences~\cite{audience_expansion}. 

Google supports different ad formats: (1) in-stream \emph{video ads} -- which appear before, during, or after a YouTube video watched on YouTube or the Display Network. The Display Network is made of applications on mobile phones or websites that partner with Google to serve their ads. A video ad can be skippable or not, depending on the parameters chosen by the advertiser. (2) \emph{floating banner ad}-- which is an image banner shown on top of the video watched for several seconds. And (3) \emph{video discovery ads}--that can appear on the YouTube search results page, alongside related videos, or on the YouTube homepage. 

\vspace{2mm}
\noindent \textbf{Transparency} All ads are marked with an "Ad" disclaimer. When users are watching an ad, they can click on  ``Why you're seeing this ad'' to learn what parameters and data were used to target the ad. Figure~\ref{fig:ad_explanaiton} (appendix) shows a screenshot of such an ad explanation.  

\subsection{Advertising on YouTube Kids} 

YouTube Kids is a video streaming platform that only contains video content curated for children~\cite{YKDef}. While YouTube Kids is widely adopted, still more parents say they are using YouTube to let their children watch videos~\cite{nscreenmedia}. 
Videos that are on YouTube Kids are curated from YouTube according to three age categories: (1) The very young: 4 years old and under; (2) The little ones: 5-8 years old; (3) The older ones 9-12 years old. 
The selection is made through a combination of algorithmic detection and manual review~\cite{YKVideo}. YouTube Kids is ``invitation only'', and creators cannot request to participate. 
Whenever a video exists on the two platforms, YouTube proposes a redirection link to YouTube Kids (see Figure~\ref{fig:youtube-kids-redirection} in appendix).

In YouTube Kids, advertising is possible, but the total number of ads delivered is limited. All YouTube Kids Paid Ads must be pre-approved by YouTube's policy team before being served in the YouTube Kids app. Ads have to follow strict creative guidelines, and many products are prohibited from being advertised, such as food and beauty products~\cite{YKAd}. 
\textit{Advertisers cannot instruct Google to show their ads on YouTube Kids}--the platform itself curates ads and whitelists them for being shown in YouTube Kids. 
Moreover, it is not possible to send interest-based ads, it is forbidden to use remarking and tracking pixels, and ads are not allowed to redirect to websites or product purchase paths~\cite{YKAd}.

\vspace{2mm}
\noindent \textbf{Transparency} Before being shown an ad, children are presented with a short intro video explaining that what comes next is an ad. In addition, the video ad is marked with an "Ad" disclaimer. On YouTube Kids, there is no equivalent to the ``Why you're seeing this ad'' button; hence, there is no way, to our knowledge, to know the precise reasons and parameters why an ad has been shown.

%!TEX root = main.tex

\section{Legislation on advertising to children}
\label{sec:bkg_law}

Historically, legislation regarding advertising and children has been developed for ads on television and not for ads online.
However, more recently, to adapt to the new online advertising ecosystem, privacy legislations such as the COPPA and online platforms legislations such as the DSA have started to include provisions regarding online advertising and children. In this section, we aggregate and review legislation related to children and advertising in the E.U. and the U.S. to understand what practices are regulated and how. Note that there is no single legal text that covers advertising to children, and the rules are generally spread across various broader legal texts. Hence, we present next the excerpts that apply to advertising and children. 
%This analysis is done by a legal scholar author and a European and consumer law expert. 
While we only interpret our findings in the context of the DSA and the COPPA, this section contains a more comprehensive set of text related to advertising to children that we hope can serve as a reference for researchers wanting to learn more about rules in the space. When possible, we cite the actual text (provided it is not too long) of the rules to not bias the description with our interpretation. However, to improve understandability, we provide an overall summary of the topics treated by the rules at the beginning of each subsection.

\subsection{Legislation on advertising on television} 
Legislation that applies to advertising to children on television imposes rules on the \emph{content} that can be advertised, on the \emph{duration} of ads, and on the \emph{separation} between ads and the television program. 

\paragraph{E.U. legislation}
In the E.U., some legal texts come from the European Union, and they have to be implemented by all the Member States, and some are set at the national level, and they apply to only a specific country. 
At E.U. level, legislation regarding advertising to children on television is contained in the larger \underline{Directive 2010/13/EU  of March 10, 2010, on Audiovisual Media} \newline \underline{Services Directive} amended by the \underline{Directive (EU) 2018/1808}''~\cite{directiveAMSD,directive2018}. 
Regarding content, the \underline{Article 9} of this directive (that applies to advertising to children) states that: ``{\em (1e) audiovisual commercial communications for alcoholic beverages shall not be aimed specifically at minors and shall not encourage immoderate consumption of such beverages.}'', and ``{\em (1g) audiovisual commercial communications shall not cause physical or moral detriment to minors. Therefore they shall not directly exhort minors to buy or hire a product or service by exploiting their inexperience or credulity, directly encourage them to persuade their parents or others to purchase the goods or services being advertised, exploit the special trust minors place in parents, teachers or other persons, or unreasonably show minors in dangerous situations.''
} 

In addition, \underline{Article 20 (2)} prohibits advertising breaks on television during children's programs of less than 30 minutes:  ``{\em The transmission of children's programmes may be interrupted by television advertising and/or teleshopping once for each scheduled period of at least 30 minutes, provided that the scheduled duration of the programme is greater than 30 minutes.}''
Finally, \underline{Article 11 (3d)}  imposes a clear separation between the advertised message and the rest of the broadcasted program: ``{\em Programmes containing product placement shall be appropriately identified at the start and the end of the programme, and when a programme resumes after an advertising break, in order to avoid any confusion on the part of the viewer.}''~\cite{decretFR}. 

This directive was transposed in every Member State of the European Union, with some national adaptations. For example, in France, the directive was transposed by a decree of 27 March 1992, amended by a decree of 2 July 2010~\cite{decretFR}.  %\update{Explain what transposition is}
%	\update{In addition to the E.U. directive, at the country level in France, legislation regarding advertising on television to children is contained in the decree of March 27, 1992.} This degree was updated in 2022 and has transposed the E.U. directive.
%In addition, at the country level in France, the 2010 European Union directive was transposed by a decree of 27 March 1992, amended by a decree of 2 July 2010 (footnote 5a)~\cite{decretFR}. 
% \underline{Article 14} of this decree imposes a clear separation between the advertised message and the rest of the broadcasted program: {\em ``The advertising messages or the sequences of advertising messages must be easily identifiable as such and clearly separated from the rest of the program, before as after their diffusion by screens recognizable by their optical and acoustic characteristics.''}. 
%In addition, \underline{Article 15} of the same decree prohibits advertising breaks on television during children's programs of less than 30 minutes~\cite{decretFR}.  

\paragraph{U.S. legislation}
%Title 47 of the Code of Federal Regulations (CFR)
The Federal Communication Commission's (FCC) rules and regulations on advertising and children are located in \underline{Title 47 of the Code of Federal Regulations (CFR)}. These regulations are colloquially referred to as the Children's Television Act (CTA). They have been initially proposed in 1990, and have been updated several times since with the latest update in 2019. 
%The bulk of the text has rules on the amount of time television programs should broadcast educational and informational programs for children. %In addition, there are also regulations on advertising in broadcast and cable television programming targeting children 12 and younger, including limits on ad time, and prohibiting the airing of advertising for products related to the program currently airing.
The regulation imposes rules on  advertising in broadcast and cable television programming targeting children 12 and younger, including limits on ad time, and prohibiting the airing of advertising for products related to the program currently airing. 
%In addition, there are regulations on advertising in broadcast and cable television programming targeting children 12 and younger.
% In October 1990, President George H. W. Bush signed the Children's Television Act (CTA), an Act of Congress ordering the FCC to implement regulations surrounding programming that serves the "educational and informational" (E/I) needs of children, as well as the amount of advertising broadcast during television programs aimed towards children
%The Federal Communication Commission's (FCC) ``commercial limits rules and policies'' applies to advertising to children.  
\underline{73.670} and \underline{76.225} state that \emph{``No cable operator shall air more than 10.5 minutes of commercial matter per hour during children's programming on weekends, or more than 12 minutes of commercial matter per hour on weekdays.''}~\cite{FCCreg1,FCCreg2}.
In addition, rules related to content in general, present in \underline{73.3999}, also apply to content advertised to children. Federal law prohibits obscene, indecent and profane content from being broadcast on the radio or T.V.~\cite{FCCreg3}.
Finally, FCC has several policies that are designed to protect children from confusion that may result from the intermixture of programs and commercial material in children's television programming in the \underline{Sponsorship Identification Rules and Embedded Advertising}~\cite{FCCreg4}.

\subsection{Legislation on advertising online}

Regulations on online advertising and children restrict some \emph{targeting mechanisms} for reaching children (namely through profiling) while they allow others (namely through contextual advertising). 
Per our research, regulations on online advertising seem to miss the precise rules present in television advertising regarding the content and the duration of ads. %As new legislations are emerging, these issues might be addressed in future acts. %Legislators are working on legal texts, and new regulations will be proposed in the following years that could address these issues. 
In addition, the E.U. legislation imposes obligations on platforms to assess \emph{systemic risk} incurred by their systems that could affect minors and an obligation to mitigate such risks. 

\paragraph{E.U. legislation}
At E.U. level, a new Regulation of 19 October 2022, the \underline{Digital Services Act} contains regulations regarding advertising to children online~\cite{DSA}. The DSA follows the European Commission Communication ``\underline{European Strategy for a Better Internet} \underline{for Children}'' regarding minors~\cite{BIK,BIK+}. At a high level, the DSA forbids advertising to children based on profiling and puts an obligation on online platforms to assess systemic risks that might harm children's rights and mental well-being and health. Interpretation hint: the actual rules are present in ``Articles'', while ``Recitals'' can provide additional explanations and intentions behind the rules.

%Minors facing digital services is a central concern of the DSA, which is itself a follow-up to a European Commission Communication entitled ``\underline{European Strategy for a Better Internet for Children}''~\cite{BIK,BIK+}. 

%Many new prohibitions (1) and obligations (2) are imposed on online platforms in the DSA, reflecting a substantial improvement in Minors protection with regard to Advertisement.
% 1. - Prohibitions for the Benefit of the Minors
%\item a) - General Prohibitions
%Dark Patterns on online interfaces, i.e. âpractices that materially distort or impair, either on purpose or in effect, the ability of recipients of the service to make autonomous and informed choices or decisionsâ , are at the heart of the European legislator's concerns, whether in the GDPR  or in the UCPD .
%The article 25 of the DSA  adds a new layer of prohibitions for online platforms providers concerning Dark Patterns that would not be regulated by the first two instruments.
%In addition to this principle of prohibition by design and of action, the Commission aims to establish âguidelines about these kind of practices, notably: 
%-	giving more prominence to certain choices when asking the recipient of the service for a decision;
%-	repeatedly requesting that the recipient of the service make a choice where that choice has already been made, especially by presenting pop-ups that interfere with the user experience;
%-	making the procedure for terminating a service more difficult than subscribing to itâ.
\underline{Article 28 (2) of the DSA} states that ``{\em Providers of \textbf{online platform shall not present advertisements on their interface based on profiling} as defined in Article 4, point (4), of Regulation (EU) 2016/679 using personal data of the recipient of the service when they are aware with \textbf{reasonable certainty} that the recipient of the service is a minor}''. 
In the \underline{Article 4 (4) of the GDPR}, ``\textbf{profiling}'' means ``{\em any form of automated processing of personal data consisting of the use of personal data to evaluate certain personal aspects relating to a natural person, in particular to analyse or predict aspects concerning that natural person's performance at work, economic situation, health, personal preferences, interests, reliability, behaviour, location or movements"}. %Appendix~\ref{app:law} presents more related nomenclature. 

The notion of ``\textbf{reasonable certainty}'' is a complex concept that can be interpreted in different ways. \underline{Article 28 (3)} states that ``compliance'' with Article 28 (2) ``{\em shall not oblige providers of online platforms to process additional personal data in order to assess whether the recipient of the service is a \textbf{minor}}''.
This paragraph is, itself, open to several interpretations, which \underline{Recital 71} clarifies as follows: ``{\em In accordance with Regulation (EU) 2016/679, notably the principle of data minimisation as provided for in Article 5(1), point (c), thereof, this prohibition should not lead the provider of the online platform to maintain, acquire or process more personal data than it already has in order to assess if the recipient of the service is a minor. Thus, this obligation should not incentivize providers of online platforms to collect the age of the recipient of the service prior to their use }''. %In view of the complexity of the wording of Article 28 (2) and (3) and of Recital 71, the Commission is encouraged, by \underline{Article 28 (1)} to issue guidelines to assist platform providers in the implementation of the duty of the providers of online platforms accessible to minors to put in place appropriate and proportionate measures to ensure a high level of privacy, safety, and security of minors, on their service. 
To comply with such rules and to not ask for additional data on the age of the users, YouTube is asking content creators to label whether their content is intended for children or not~\cite{Youtube_kids_coppa,Youtube_kids_coppa2}.

% 2. - From Due Diligence Obligations to Admistrative Monitoring and Civil Liability
%In Chapter III of the DSA (Due Diligence Obligations for Very Large Online Platforms and Search Engines). 
Furthermore, given the importance of very large online platforms and search engines (VLOPS), in facilitating public debate, economic transactions, and the dissemination to the public of information, opinions, and ideas and in influencing how recipients obtain and communicate information online, the \underline{Section V of Chapter III} \underline{of the DSA} (Due Diligence Obligations for Very Large Online Platforms and Search Engine)
 impose specific obligations on the providers of very large online platforms in addition to the obligations applicable to all online platforms. 
In particular, \underline{Articles 34 and 35 of the DSA} place an obligation on VLOPS to assess four broad categories of {\em systemic risk}, as well as an obligation to mitigate such risks.  VLOPS must therefore be particularly vigilant about these risks, which involve the impact of their recommendation, moderation, and advertising systems on the health, safety, and fundamental rights of consumers, whether minors or adults. 
The two systemic risks that relate to advertising and children are described in \underline{Recital 81} and \underline{Recital 83}:
%\underline{Recital 81}, explaining those Articles, states that an important category of risks  ``{\em concerns the actual or foreseeable impact of the service on the exercise of fundamental rights, as protected by the Charter, including but not limited to human dignity, freedom of expression and of information, including media freedom and pluralism, the right to private life, data protection, the right to non-discrimination, the \textbf{rights of the child} and consumer protection. Such risks may arise, for example, in relation to the design of the algorithmic systems used by the very large online platform or by the very large online search engine or the misuse of their service through the submission of abusive notices or other methods for silencing speech or hampering competition. When assessing risks to \textbf{the rights of the child}, providers of very large online platforms and of very large online search engines should consider for example how easy it is for \textbf{minors} to understand the design and functioning of the service, as well as how minors can be exposed through their service to content that may impair minors' health, physical, mental and moral development. Such risks may arise, for example, in relation to the design of online interfaces which intentionally or unintentionally exploit the weaknesses and inexperience of \textbf{minors} or which may cause addictive behavior}''.

\underline{Recital 81} ``{\em concerns the actual or foreseeable impact of the service on the exercise of fundamental rights, as protected by the Charter, including but not limited to human dignity, freedom of expression and of information, including media freedom and pluralism, the right to private life, data protection, the right to non-discrimination, the \textbf{rights of the child} and consumer protection. Such risks may arise, for example, in relation to the design of the algorithmic systems used by the very large online platform or by the very large online search engine or the misuse of their service through the submission of abusive notices or other methods for silencing speech or hampering competition. When assessing risks to \textbf{the rights of the child}, providers of very large online platforms and of very large online search engines should consider for example how easy it is for \textbf{minors} to understand the design and functioning of the service, as well as how \textbf{minors} can be exposed through their service to content that may impair \textbf{minors}' health, physical, mental and moral development. Such risks may arise, for example, in relation to the design of online interfaces which intentionally or unintentionally exploit the weaknesses and inexperience of \textbf{minors} or which may cause addictive behavior}''.

\underline{Recital 83} identifies another important risk:  ``{\em category of risks stems from similar concerns relating to the design, functioning or use, including through manipulation, of very large online platforms and of very large online search engines with an actual or foreseeable negative effect on the protection of public health, \textbf{minors} and serious negative consequences to a person's physical and mental well-being, or on gender-based violence. Such risks may also stem from coordinated disinformation campaigns related to public health, or from online interface design that may stimulate behavioural addictions of recipients of the service}''.

Finally, regarding the content of online ads, we found a broad provision in the \underline{Directive (EU) 2018/1808 of 14 November 2018} amended by \underline{Directive 2010/13/EU} (Audiovisual Media Services Directive). The amended directive states in \underline{Article 28b (a)} that ``{\em Member States shall ensure that video-sharing platform providers under their jurisdiction take \textbf{appropriate measures to protect minors} from programmes, user-generated videos and \textbf{audiovisual commercial communications which may impair their physical, mental or moral development}}''.  Where the appropriate measures are strict access control, age verification and parental control (see Appendix~\ref{app:law} for the description of appropriate measures). %Where \underline{Article 6a(1)} states that : ``{\em Member States shall take appropriate measures to ensure that audiovisual media services provided by media service providers under their jurisdiction which may impair the physical, mental or moral development of minors are only made available in such a way as to ensure that minors will not normally hear or see them. Such measures may include selecting the time of the broadcast, age verification tools or other technical measures. They shall be proportionate to the potential harm of the programme. The most harmful content, such as gratuitous violence and pornography, shall be subject to the strictest measures}''.
Note that, this article applies to all continent that might be seen by minors on video-sharing platforms and is not specific to advertising. 

\paragraph{U.S. legislation}
The \underline{Children's Online Privacy Protection Rule} \newline (COPPA)~\cite{coppa} has restrictions on advertising to children. The proposed rules are intended to stop the behavioral advertising, retargeting, and profiling of children under 13, but not contextual advertising. 
More precisely, COPPA states that advertisers and content owners may not collect any personal information (which includes cookies and other persistent identifiers) from children under 13 years of age without \textbf{verifiable parental consent} (see Appendix~\ref{app:law} for a definition of verifiable legal consent and personal infromation). 
However, COPPA allows operators to collect \textbf{personal information} only for ``{\em support for the internal operations of the Web site or online service means that those are activities necessary to (...) authenticate users of, or personalize the content on, the Web site or online service (...) \textbf{serve contextual advertising} on the Web site or online service or cap the frequency of advertising (...) so long as the information collected for these activities is not used or disclosed to contact a specific individual, including through behavioral advertising, to amass a profile on a specific individual, or for any other purpose'''}.

Senators introduced in January 2022 in front of the US Congress a proposal for ``\underline{Banning Surveillance Advertising Act}''~\cite{BSAA}. This text, which is at the proposal stage, intends to prohibit targeted advertising, under certain conditions, but regardless of the age of the recipient, whether a minor or an adult. 
%\oana{what do they propose; anything specific to children?} \oana{do you also have a reference for the similar voting in the european parliment.}

%\oana{is there anything about content?}

%precise rules on the content of online ads as well as the 
%\oana{not a lot of regulation on the contnet in online advertising and the duration of ads}
%The \underline{Recital 94} specifies that ``{\em the obligations on assessment and mitigation of risks should trigger, on a case-by-case basis, the need for providers of very large online platforms and of very large online search engines to assess and, where necessary, adjust the design of their recommender systems, for example by taking measures to prevent or minimise biases that lead to the discrimination of persons in vulnerable situations, in particular where such adjustment is in accordance with data protection law and when the information is personalised on the basis of special categories of personal data referred to in Article 9 of the Regulation (EU) 2016/679}''. 

%AJOUT 

%!TEX root = main.tex

\section{Mechanisms for targeting children}
\label{sec:targeting}

When advertising, Google does not allow advertisers to choose audiences of users under 18. Hence, the advertising interface does not allow advertisers to target children directly.  
However, placement-based advertising can be easily exploited to target children with ads: \textit{an advertiser simply needs to aggregate a list of videos or channels with content for children and target its ads at these videos.} 
Next, we evaluate whether such targeting works in practice. We first describe how one can aggregate a list of children-focused videos. We then describe how we performed ad experiments to test whether Google allows placement-based advertising on children-focused videos.

\subsection{Curated children-focused videos}

We compiled two lists of children-focused videos and one list of general-audience videos (as control): 

\vspace{1mm}
\noindent \KidsList--To compile a list of children-focused videos, we first take advantage of YouTube Kids and, hence, YouTube's own characterization of videos as being dedicated to children. For this, we created an adult account, and we used it to create three children's accounts corresponding to the three age groups proposed by YouTube Kids (under 4, 5 to 8, and 9 to 12). We then manually browse the different categories of videos recommended when navigating to the broadcasts, music, learning, and discovery tabs in YouTube Kids and collect the corresponding channels. 
We collected \emph{22 YouTube Kids channels}. 
For each channel, we then collect the ten most recent videos from its YouTube Kids channel page.

\vspace{1mm}
\noindent \SeedList--We observed that not all children-focused videos available on YouTube are part of YouTube Kids. To build a complementary list, we took three channels with a large number of subscribers: Swan \& N\'eo (5.97~M subscribers), Madame R\'ecr\'e FR (3.81~M subscribers), and Studio Bubble Tea (1.82~M subscribers), and did a snowball sample on the ``Channels'' section. The ``Channels'' section contains channels with similar content according to YouTube or that have been manually added by the channel's owner.  
We added the channels collected (avoiding duplicates) at the end of our list and browsed this list in FIFO. Our stopping point was to have 200 YouTube channels in total. We then manually filtered out channels present on YouTube Kids (one can copy-paste the YouTube link on YouTube Kids to check if it exists) and channels not mainly intended for children according to the FTC guidelines~\cite{FTC-content}. Finally, we kept a list of \emph{20 children-focused channels}, and we collected the ten most recent videos from their YouTube channel pages.

\vspace{1mm}
\noindent \ControlList--We compiled a list of general-audience channels to have a point of reference. For this, we created a profile with no search history or preferences. We collected the channels corresponding to videos recommended on YouTube when logging in for the first time. We compiled a list of \emph{20 adult-focused channels} (the intersection with the children-focused lists is empty), and we collected the ten most recent videos from their YouTube channel pages. 
%Our channel lists can be found at \update{\emph{anonymized}}. 
The lists we compiled are not representative and are biased toward videos recommended by YouTube and more popular videos on the platform. Nevertheless, representativeness is not necessary if the goal of the ad experiments is just to test whether something is possible. 

\begin{table*}[htp]
\caption{Targeting parameters of our ad campaigns and their impressions. * Total impressions (out of which impressions on videos in \KidsList; and in \SeedList).} 
 \vspace{-3mm}
\begin{center}
  \footnotesize{
\begin{tabular}{p{0.5cm}|p{6.5cm}|p{6cm}|p{1cm}|l}
\toprule
\textbf{CID}  & \textbf{Placement} & \textbf{Interests} & \textbf{Location} & \textbf{Impressions*} \\ 
\midrule
1 & 3  \KidsList videos & None & U.S. & 450 (450; 0) \\ 

 2   & 4 \ControlList videos + 7 \KidsList videos + 1  \SeedList video & None & Fr. & 605 (43; 75)  \\%Con - 04 - 11 - 2022
 3  & 10 \ControlList videos + 10  \KidsList videos + 10 \SeedList videos & None & Fr.  & 6,124 (1,414; 2)\\%Vidéo Considération - 2022-06-17%
 \arrayrulecolor{gray}\specialrule{0.3pt}{1pt}{1pt} 
4  & 3 \ControlList videos + 6 \KidsList videos + 1 \SeedList video & News \& Politics, Travel, Sports \& Fitness & Fr. & 941 (121; 471)\\  % Con Int - 04-11-2022
5  & 20 \KidsList channels + 4 \SeedList channels & Shoppers,  Sports Game Fans & U.S. & 10,854 (1,657; 9,203) \\  % Oana Video Efficient reach - US Kids - 2022-11-24
% 6 & Placement + interests &  20 \KidsList channels  &  Shoppers,  Sports Game Fan & \\
6  &  20 \KidsList videos  &  Shoppers, Food \& Dining, Education, Media \& Entertainment, Beauty \& Wellness, Banking \& Finance & U.S. & 4,650 (1,058; 0) \\ %  Oana Efficient Reach  2023-01-31 - Kids videos  ad group video efficient reach 
7  &  20 \KidsList videos  &  Shoppers, Food \& Dining, Education, Media \& Entertainment, Beauty \& Wellness, Banking \& Finance & E.U.$\dagger$& 2,372 (2,372; 0)\\%  Oana Efficient Reach  2023-01-31 - Kids videos  ad group video efficient reach #2
%Total on children: 13436
 \arrayrulecolor{black} \bottomrule
\end{tabular}
}
\footnotesize{ $\dagger$ Belgium, Denmark, Finland, France, Germany, Italy, Netherlands, Poland, Romania, Spain, Sweden.}

\end{center}
\label{tab:tae}
\vspace{-5mm}
\end{table*}%

\subsection{Targeting ad experiments}
\label{sec:ad_targeting}
To check whether Google allows advertisers to target their ads to children-focused videos, we created \emph{seven} real-world ad campaigns in which we instructed Google to place our ads on children-focused videos.  Section~\ref{sec:ethics} describes how we chose the content of the ads to minimize risks for children. 
 As campaign parameters we chose the ``Brand Awareness and Reach'' campaign objective, instructed Google only to show the ad on YouTube (and not on its Display Network), and disabled the ``audience expansion'' option. We left all other parameters on the default value.  We launched three ad campaigns where we only used placement-based targeting (CID 1 to 3 in Table~\ref{tab:tae}) and four ad campaigns combining both placement-based and interest-based targeting (CID 4 to 7 in Table~\ref{tab:tae}).
 For each ad campaign, we chose various combinations of placements: videos or channels from \KidsList, \SeedList, and \ControlList. CID 1 and 5-7 only use children-focused videos as placement, while CID 2-4 combine adult-focused with children-focused videos as placement.  Our ad experiments were launched between Sept. and Nov. 2022. \textbf{All our seven ad campaigns have been validated by Google and delivered to users.}  Nevertheless, this information alone is not enough to prove that Google delivered our ads on children-focused videos, as the platform could have potentially not respected our criteria and delivered our ads on videos that are not on our list.

Luckily, Google provides advertisers with an interface where they can check statistics about the delivery of their ad campaigns and allows us to check for each ad campaign the total number of ad impressions as well as the number of ad impressions on each video the ad was placed on~\cite{GoogleAds}. %We checked for each ad campaign the total number of ad impressions and the number of ad impressions on each specific video list.  
Table~\ref{tab:tae} reports the total number of impressions as well as impressions on \KidsList and \SeedList videos only.  
The table shows that all our ad campaigns resulted in our ads  being shown on children-focused videos including videos sourced from YouTube Kids--these videos are marked by the platform itself as being primarily intended for children.  
%This information provides evidence that our  all our ad campaigns resulted in our ads being shown on children-focused videos. 
%Hence, this shows \textbf{it is technically possible to target children with ads through placement-based targeting on YouTube}.
Hence, this shows that \textbf{advertisers can target children with ads on YouTube by asking the platform to place their ads on children-focused videos}. %Note that we do not know whether our ads were shown on YouTube or YouTube Kids since it is the platform that decides.

Placement-based targeting is done independently of the identity of the connected user and their Google profile. 
Hence, we performed four experiments where we instructed Google to combine placement-based with interest-based targeting in order to target ads both based on the video watched and based on the profile of the user watching the video.
We perform this experiment because both the DSA and COPPA ban or restrict the use of profiling in advertising to children.
Table~\ref{tab:tae} (CID~4~to~7) shows the placements and the interests we used.   To check if Google has a different treatment for videos on YouTube Kids (as the platform knows they are intended for children), for CID~6 and CID~7, we used as placement only videos on YouTube Kids. 
We also checked if Google has different policies across different countries, maybe due to differences in legislation. CID~6 was targeted at the U.S., while CID~7 was targeted at 11 countries in the European Union. 
All four ad campaigns were validated and shown to users. Again, this is not enough to prove that Google is serving ads based on the profiles of users on children-focused videos, as Google might have ignored the profiling part of our request. 
 %We again double-checked, using the statistics interface provided by Google, that our ads were placed on the videos we asked for. %All ad campaigns besides CID~5  were only shown on the videos we requested. CID~5 was shown on 1,510 other placements. We do not know why Google showed them in other places. 
Luckily, Google also provides statistics about the number of impressions shown to users with particular characteristics. For example, the statistics provided by Google for CID~4 say it had 393 impressions on users interested in ``Travel'', 387 in ``Sports \& Fitness'' and 161 in ``News \& Politics''. 
The statistics provided by Google indicate that for all four ad campaigns (CID~4~to~7), Google delivered the ads to only users with the interests we asked for (i.e., the total number of impressions across placements = the total number of impressions across interests). 
Hence, \textbf{Google allows advertisers to perform targeting based on profiling when placing their ads on children-focused videos.} Such a combination of targeting is allowed both in the U.S. and E.U.; it is also allowed on videos marked by YouTube itself as intended for children. 
 
%CID 4 
%Travel 393
%Sports \& Fitness 387 
%News \& Politics 161
%
%CID 5
%Shoppers 9,346
%Media \& Entertainment  1
%Other 1,510
%
%CID 6
%Shoppers 3,615
%Food \& Dining 1,035
%
%CID 7
%2,372
%Food \& Dining 2,328
%Shoppers 44

\subsection{Ethics statement}
\label{sec:ethics} 

We ran ad experiments on real ad platforms and targeted children-focused videos with our ads. Hence, our ads have probably been seen by children.  
We chose the content of the video ad and the landing page to limit potentially harmful effects on children. We first created a short video that records trees while walking in a park (the video does not capture any visible human face). Second, we used an already existing children-focused video on YouTube under a Creative Commons License that allows redistribution (Figure~\ref{fig:ad1}, in the appendix, show a screenshot of our two ads). Finally, the landing URL of the ad was either the tourism office of the city we lived in or the YouTube Channel page corresponding to the video we used as an ad. 
We set small ad budgets to limit the number of children that see our ads and stopped the ad campaign as soon as we saw impressions. We did an informal validation with colleagues (including legal scholars, behavioral economists, and cognitive scientists), and they did not express any concern regarding potential harm from our video ads. 

We filed a request with our institute's DPO (Data Protection Officer) for the data collection in Section~\ref{sec:ad_targeting}, and we were exempt as we do not collect personal information from users in these experiments. The data collection gathered through a browser extension (Section~\ref{sec:metho}) and the ad experiments to assess risks with online platforms (Section~\ref{sec:ad_targeting}) were covered by a project-level IRB.

\subsection{Implications}  

Our experiments show that \emph{is possible} for advertisers to directly market to children on YouTube through placement-based targeting on children-focused videos. In light of the texts applicable in the E.U. and the U.S., it appears that the online targeting of children by contextual advertising \emph{is permitted}. Hence, placement-based advertising is also permitted as it is a form of contextual advertising. 
Nevertheless, contextual advertising targeting children online is less secure than contextual advertising targeting children on television. Online advertisers can place an ad under a single piece of content (e.g., a particular cartoon or a particular video clip dedicated to children), which can make verification of the content of the ad very difficult in practice, as it will only be seen by children who have seen that content, with \emph{very little possibility of effective verification} of the content of the ad by adults.

In contrast, the online targeting of children by profiling (such as interest, location, and behavior) is clearly banned in the European (DSA) and \emph{might} be restricted in the American (COPPA). In fact, COPPA restricts the \emph{collection} of personal information on content primarily intended to children, however, it is not clear from the text if it also restricts the \emph{use} of previously complied user profiles (e.g., profiles complied when the users was browsing adult-focused videos) when serving an ad on a children-focused video.  
 Hence, it appears from the experiments that there is a need to better distinguish between cases where targeting is carried out solely on the basis of context or where targeting is carried out by mixing contextual placement and profiling. This distinction is necessary to effectively implement the  DSA and  COPPA texts protecting children against advertising based on profiling.

Nevertheless, in 2019, the Federal Trade Commission (FTC) served YouTube with a \$170 million fine for violating COPPA because it had illegally collected information on children for its personalized ads. To avoid future fines, YouTube committed to better distinguish content intended primarily for children by relying on a combination of self-reports from creators and algorithms. Since then, content creators have been obligated to label their videos as ``Made for Kids'' (or risks \$45k fines otherwise). \emph{YouTube said they would assume any viewer of children-focused videos is underage, no personal data collection will be performed on such content, and no personalized ads will be sent on such content} (as per the COPPA rule)~\cite{washingtonpost_coppa, Youtube_kids_coppa,Youtube_kids_coppa2,FTC_COPPA_YouTube_set}.
Hence, contrary to this declaration, we were able to send personalized ads on such content.

%{(2)} Furthermore, it appears from the experiments that there is a need to better distinguish between cases where targeting is carried out solely on the basis of context or where targeting is carried out by mixing contextual placement and profiling. This distinction is necessary to effectively implement the European (DSA) and American (COPPA) texts protecting children against advertising based on profiling.

%!TEX root = main.tex

\section{Usage of placement-based targeting}
\label{sec:metho}

This section proposes a measurement methodology to investigate whether advertisers, in the wild, are exploiting placement-based targeting to advertise their products to children.

\subsection{Data collection methodology}
Our methodology is based on a browser extension that can capture information about the ads shown when a user is watching a YouTube video. 
At a high level, the measurement methodology consists in instructing six browser profiles to watch children-focused videos and collect the ads shown. 

\paragraph{Ad capture tool}

Each time an ad is shown to the user, the DOM is updated to add new elements to the HTML code of the video watched. So, to capture ads, our browser extension simply monitors the DOM for changes. 
Our extension works on Chrome and can capture both \emph{video ads}--video ads launched at the beginning, middle, or end of a watched video; and \emph{float card ads}--ads containing an image, a GIF, or text showed while watching a video above the navigation bar. When our browser extension detects an ad, it collects the following information (mainly textual): 

 \hspace{1mm} $\bullet$ \textit{Ad data}: 
The name of the advertiser, the advertiser's YouTube channel, the advertiser's link, the image or GIF in the case of float ads, and the URL of the ad for video ads. 
     
\hspace{1mm} $\bullet$ \textit{Video watched data}: 
    We collect the title of the video watched (the video on which the ad appears), its URL,  its description,  the nb. of views, the date the video was published, the corresponding channel, and the nb. of channel followers. 
   
\hspace{1mm} $\bullet$ \textit{YouTube Kids redirection}: We collect if YouTube proposes redirecting the user to YouTube Kids (it means the video is on YouTube Kids--see Figure~\ref{fig:youtube-kids-redirection} in appendix). 
 
\hspace{1mm} $\bullet$ \textit{Ad explanation}: 
    The browser extension collects data that appears in the ``Why am I seeing this ad?'' feature. This feature can be accessed by clicking on the button  \textcircled{\textit{i}} at the bottom left of the ad video. A modal opens and describes to users the reasons the ad was shown to them. We collect this information by simulating a click on the button, and thus we recover the response to the request sent after the click. To close the modal, we simulate another click outside the modal's window.

We do not collect information about the profile/identity watching the video (each user is associated with a unique identifier generated by the browser extension). This data is sent and stored in our lab servers through a secure HTTPS connection. The storage server is protected by a firewall.

\paragraph{Personas}
The ads shown when someone watches a video on YouTube can depend on the browser profile (as determined by the cookies present in the browser that track sites previously visited) and the Google profile of the user in case the user is logged in. Hence, to have a broader view and account for how the browser profile could influence the results, we created six browsing profiles with different parameters--see Table~\ref{tab:profiles}. 

We created: (1) Two browser profiles without a Google profile associated and no browsing history. (2) Two profiles are associated with an empty Google profile with no gender specification and no browsing history (the profiles only have the age set because the users are obliged to specify the date of birth to create a Google account). 
And (3) two profiles are associated with Google accounts and browsing history. One account has the gender set to female and the other to male. For the female profile, we visited four sites related to clothes, cats' food, Volkswagen cars, and animal rights in France; for the male profile, we visited one site related to plastic surgery. 
For P3-P6, we checked the Google Ads Settings page to investigate what interests Google has inferred for the different profiles~\cite{adsettings}. P3 and P4 have, as expected, no interests. The P5 profile had Fashion \& Trend, Pets, and Auto \& Vehicles as interests, and P6 had Body Care \& Fitness as interests. These interests are consistent with the sites consulted. 
 
Google does not allow children (younger than 18 years old) to create Google profiles. Our profile choice is motivated by the possibilities offered by Google for how a child could watch videos on YouTube: (1) a child can use an incognito browser that is not connected to any profile;  (2) a child could create an account with an erroneous age; or (3) a child can use the account of one of his/her parents, who may have already used this account on sites and applications so that browsing history is already present.

\begin{table}
  \caption{Characteristics of our browser profiles. }
 \vspace{-3mm}
  \label{tab:profiles}
  \begin{center}
\footnotesize{
\begin{tabular}{cp{7cm}}
\toprule
\textbf{Profile}  & \textbf{Details} \\ \midrule
P1 & No Google profile + no search history \\
P2 & No Google profile + no search history \\
P3 & Empty google Profile + no search history and no gender\\
P4 & Empty google Profile + no search history and no gender\\
P5 & Google profile (woman)  + browse four sites related to clothes, cats food, Volkswagen cars and animal's rights in France  \\
P6 & Google profile (man) +  browse one site related to plastic surgery  \\
\bottomrule
\end{tabular}
}
\end{center}
\vspace{-5mm}
\end{table}

\paragraph{Instrumentation}
We conducted our data collection on YouTube between April 28th 2022 and May 5th 2022.
We collected data simultaneously from the six profiles on two different computers. We launch the six personas simultaneously by creating six Selenium sessions. The personas browse the ten most recent videos in each of the channels in the three lists: \KidsList, \SeedList, and \ControlList. Each video was watched from start to finish. With the help of the extension, each ad served is collected and stored on our servers. Our stopping point is the end of the entire YouTube channel list--620 videos.
Table~\ref{tab:ads} shows the number of video and float card ads we collected for each list and each profile. Across the six profiles, we collected 3,221 ads. 

\noindent \textit{\underline{Note on experimental difficulties:}} The data collection is time-consuming:  (1) we need to watch videos until the end to catch all ads because ads can appear at random times during the video, and there is no fixed number of ads that appears during a video. The length of the videos varies from less than 2 minutes to more than one hour. 
And (2) our simulated double-click to recover the information in the ``Why am I seeing this ad?'' sometimes triggers a modal that does not close, interrupting the video's progress and therefore requiring human intervention to close the modal and restart the video. This limits the scale of the experiment. Nevertheless, since this paper focuses on whether advertisers use placement-based advertising to target children and not on providing a representative characterization of ads shown on children-focused videos, we believe that our instrumentation with six browser profiles is enough.

\begin{table}
  \caption{Number of ads received by the different profiles. }
   \vspace{-3mm}
  \label{tab:ads}
  \begin{center}
\footnotesize{
\begin{tabular}{ccc}
\toprule
\textbf{Profile}  & \textbf{No. of Video ads} & \textbf{No. of Float ads}  \\ \midrule
P1 & 239 & 0 \\
P2 & 205 & 0\\
P3 & 437 & 162\\
P4 & 396 & 139\\
P5 & 541 & 393 \\
P6 & 441 & 368  \\
 \arrayrulecolor{gray}\specialrule{0.3pt}{1pt}{1pt}  \arrayrulecolor{black}
ALL & 2,259 & 1,062 \\
\bottomrule
\end{tabular}
}
\end{center}
%\vspace{-6mm}
\end{table}

\subsection{Detection of placement-based targeting}
\label{sec:pl}

The reasons why an ad is shown on a video depend on the targeting parameters set by the advertiser and the platforms' algorithms to deliver ads. In this paper, we focus on placement-based advertising. 
Hence, given a set of ads shown on a video, the challenge is to distinguish between ads targeted at the video (i.e., placement-based) and ads that were shown due to other reasons (e.g., interest-based, re-targeting).

Our first idea was to watch the same video simultaneously, using multiple browser profiles, and consider only ads present in all profiles as placement-based ads. Out of the 668 unique ads we collected across the six profiles, only two ads were received by all six profiles. For information, 39 ads were received by four or more profiles, and 128 ads were received by two or more profiles. 
Nevertheless, this method sadly suffers from both false positives (the ad appears in all profiles because the advertiser launched a very big campaign and it is outbidding the competition) as well as false negatives (an ad might be placement-based but does not appear in all videos because of budget constraints and competition). Therefore, this method cannot provide reliable ground truth. 

Luckily, Google provides users explanations for why they received a particular ad in the ``Why you're seeing this ad'' feature (see Section~\ref{sec:background}). Hence, we decided to investigate whether these explanations are precise enough to help us distinguish between placement-based ads and the rest.  
Our Chrome extension was able to collect the information in the   ``Why you're seeing this ad'' button for all the ads we collected. 
Table~\ref{tab:reasons} shows the different explanations provided by Google. Note that our study was conducted in French, so the reasons in the table are translated from French and might have a slightly different syntax in English. We group these reasons into eight broad targeting categories: P--placement-based ads; R--Re-targeting ads; T--Theme-based ads; K--Keyword-based; I--Identity-based; B--Behavioral or interest-based; L--Location-based; and G-General targeting. The different ad explanations cover the different targeting types offered by Google, and the explanations provided are quite precise such as ``The topics on the website you were looking at'' or ``Estimate of your housing occupancy status by Google''. More interesting for us, the explanation with EID~1, ``The video you are watching'' seems to correspond to placement-based advertising. Another interesting explanation is EID~27, ``Ad personalizing is disabled for this account or content. Therefore, this ad is not personalized based on your data.'' as it clearly states that the ad is not personalized. This is interesting because Google claims to not deliver personalized ads on children-focused videos~\cite{Youtube_kids_coppa,Youtube_kids_coppa2}.

\begin{table}[h!]
  \caption{List of the explanations why a user receives an ad. We group them into eight targeting categories: A--placement-based ads; R--Re-targeting ads; T--Theme-based ads; K--Keyword-based; D--Demographic-based; B--Behavioral or interest-based; L--Location-based; G--General targeting. The grouping is made based on our intuition and not based on ad experiments with definite proof.}
  \vspace{-3mm}
 % select ad_channel_description, host_channel_name, title,  count(*)  as c  from videoads_aggregated inner join StoreToDB_adreason as r on videoads_aggregated.adreason_id=r.id where user_profile in (1,2,3,4,5,6) group by adreason_id  order by c  ;
  \label{tab:reasons}
  \begin{center}
  \footnotesize{
\begin{tabular}{p{0.6cm}p{0.6cm}p{6.5cm}}
\toprule
\textbf{EID} & \textbf{Cat.} &\textbf{``Why am I seeing this ad?'' explanations} \\ \midrule
1 & P&\textbf{The video you are watching.} \\
 \arrayrulecolor{gray}\specialrule{0.3pt}{1pt}{1pt}
2 &K & Your current search terms.  \\
 \arrayrulecolor{gray}\specialrule{0.3pt}{1pt}{1pt}
3 & T&The topics on the website you were looking at. \\
4 & T&Popular products from this advertiser.  \\
 \arrayrulecolor{gray}\specialrule{0.3pt}{1pt}{1pt}
5 & B& Your activity while signed in to Google. \\
6 & B&Your estimated interests based on your activity on Google (e.g. searches) on this device.\\
7 & B&Google's estimate of your interests based on your activity (e.g. searches) while you were signed in to Google.       \\
8 & B&Your activity, while you were signed in to Google.  \\
9 & B&Your activity on this device.       \\
 \arrayrulecolor{gray}\specialrule{0.3pt}{1pt}{1pt}
10 & D&Your gender.  \\
11 & D&Estimate of your level of education by Google.\\
12 & D&The age you added to your Google Account. \\
13 & D&Estimate of your housing occupancy status by Google. \\
14 & D&Your gender estimated by Google based on your activity on this device. \\
15 & D&Estimation of your parental status by Google.\\
16 & D&Your age group.  \\
 \arrayrulecolor{gray}\specialrule{0.3pt}{1pt}{1pt}
17 & R & Your visit to the advertiser's website or app.  \\
18 & R&The websites you visited.       \\
19 & R&Websites you've visited.       \\
20 & R-L&The time of day, the website you were viewing or your general location (for example country or city).\\
21 & R & Your similarities to the groups of people the advertiser is trying to reach.\\
 \arrayrulecolor{gray}\specialrule{0.3pt}{1pt}{1pt}
22 &L & Your geographic location (your country or city, for example).\\
23 & L&The time of day or your general geographic location (e.g. country or city). \\
24 & L&Google's estimate of your current approximate location.       \\
25 & L&Google's estimate (based on your past activities) of general geographies you may be interested in.       \\
 \arrayrulecolor{gray}\specialrule{0.3pt}{1pt}{1pt}
26 & G&Time of day.       \\
27 & G & \textbf{Ad personalizing is disabled for this account or content. Therefore, this ad is not personalized based on your data. Its distribution depends on other factors (such as the time or your geographical position).} \\

 \arrayrulecolor{black} \bottomrule
\end{tabular}
}
\end{center}
%\vspace{-6mm}
\end{table}

\paragraph{Validation} 
Previous works have shown that ad explanations on Facebook were incomplete and misleading~\cite{andreou2018investigating}. While it is outside the scope of this paper to perform a full audit of ad explanations on YouTube, we validate the correspondence between placement-based targeting and the ``The video you are watching? ad explanation. For this, we analyze its: (1) \textit{sensitivity} -- the extent to which when an advertiser uses placement-based targeting, this is the corresponding ad explanation Google provides users; and its (2) \textit{specificity} -- the extent to which this ad explanation does NOT appear on ads that do not use placement-based targeting. For example, if the ad explanation EID~1 also appears on ads that use theme-based targeting, then it is not specific to placement-based targeting.  

We created a Google ad campaign targeting only one specific video to investigate the sensitivity. Then we instructed the six browser profiles to watch the video we targeted, collect whether our ad was shown, and collect the corresponding explanation from Google. 
In every instance where we collected our ad, Google provided the ``The video you are watching'' (EID~1) ad explanation. We repeated the experiment in three ad campaigns with other placements and always got EID~1. Hence, as expected, EID~1 is given for ads that use placement-based targeting. %This proves the sensitivity of EID~1 with respect to placement-based targeting.

The specificity is harder to demonstrate as the ad platform provides numerous targeting parameters to advertisers, and it is hard to prove that no other combination of targeting parameters results in an EID~1 explanation. The difficulty comes from a combination of technical and financial issues: it is hard using a few browser profiles to catch an ad that was, for example, targeted at users interested in ``Shopping'' when being on a limited ad budget, as the likelihood of our ad being seen by our profiles is very low. 
%
%
%One important limitation of our study is that we rely on ad explanations provided by Google to detect the corresponding targeting parameters used by advertisers. While we tested the correctness of some of these ad explanations (in particular for placement-based advertising), we could not test all possible combinations of targeting parameters and check the corresponding explanations.  
%This means that the fraction of ads targeted using placement-based targeting might be lower if the ad explanation ``The video you are watching?. is also given for other types of targeting. This would be hard to prove as 
%Proving specificity\footnote{Specificity means that if we receive the ad explanation ?The video you are watching?, it always corresponds to placement-based targeting and not to other types of targeting.} is hard as even if we could manage to test all the different combinations of targeting parameters offered by ad platforms (which is financially prohibitive in our case), there could still be a theoretical chance, there are parameters outside our control that could trigger this ad explanation. 

Hence, we tested the specificity with respect to four types of targeting (theme-based, behavioral-based, demographic-based, and general-based), which is a subset of all possible targeting strategies, but covers some of the most used targeting techniques. For this, we again created four ad campaigns that used the four types of targeting and tried to collect our ads and check the explanation provided by Google. For none of the ad campaigns, we collected a ``The video you are watching'' explanation, and the ad explanations provided were coherent with our categorization in Table~\ref{tab:reasons}. 
Hence, while we cannot prove absolute specificity, we show the specificity of EID~1 with respect to these four categories of targeting. 

%Nevertheless, we performed several other ad campaigns with different targeting parameters (theme-based, behavioral-based, demographic-based, and general-based) to check the corresponding ad explanations. We were able to collect the ad explanations for four ad campaigns, and none of them had EID~1. 

One concern we did not manage to test is what explanation Google provides if an advertiser selects the ``audience expansion'' feature when using placement-based targeting. This feature allows Google to deliver the ad on videos other than the ones initially specified by the advertiser. If Google provides the EID~1 explanation for all the ads it delivers as part of this campaign, this could lead to false positives, as the EID~1 explanation will appear on videos selected by Google in addition to the videos specified by the advertiser in its placements.

 %is that we do not know, and we were not able to design an ad experiment that collects the ad explanation provided by Google in case an advertiser selects the ``audience expansion'' feature when using placement-based targeting--meaning the advertiser might have initially targeted its ad to certain videos, but Google is also showing the ad on other videos it deems relevant. Hence, this could lead to false positives in the sense that the advertisers might not have targeted the children-focused videos we monitor but other videos. 

\vspace{2mm}
\noindent \underline{Limitation:} Since we cannot fully prove the specificity of ``The video you are watching'' ad explanation, our results should be read as an upper bound on placement-based targeting.

\vspace{2mm}
\noindent {\underline{Note:} We conducted our data collection on YouTube rather than YouTube Kids. The main technical reason behind this is that YouTube Kids does not offer users the ``Why am I seeing this ad?'' feature; hence, we cannot deduct when advertisers use placement-based advertising. 
Nevertheless, surveys claim that YouTube is still used more than YouTube Kids by parents~\cite{nscreenmedia}. Hence, it also makes practical sense to investigate placement-based advertising on YouTube.

%!TEX root = main.tex

\begin{table}[t]
  \centering 
    \footnotesize{
      \caption{Overlap between video ads across different profiles.}
      \vspace{-3mm}
  \begin{tabular}{c|c|c|c|c|c}
\toprule
&  \textbf{P2} & \textbf{P3} & \textbf{P4} & \textbf{P5} & \textbf{P6}  \\ 
\midrule

\textbf{P1}	& 	\cellcolor{color2} 19\%  & 	6.8\%	& 8.3\%	& 8.60\%& 	14\%\\
\textbf{P2}	& 		& 2.5\%	& 2.9\%& 	2.8\%& 	3.7\%\\
\textbf{P3}	&&			&  \cellcolor{color2} 19\%	&  \cellcolor{color2} 20.5\%	&  \cellcolor{color2} 16.7\%\\
\textbf{P4}	&&&			& 	 \cellcolor{color2} 16.2\%	& \cellcolor{color2} 14.3\%\\
\textbf{P5}	&&&&				&  \cellcolor{color2} 23\% \\
\bottomrule

  \end{tabular}

  \label{tab:overlap}
  }
\end{table}

\begin{table*}[t]
  \centering 
    \footnotesize{
      \caption{Targeting categories (sourced from ad explanations) for ads received by the six profiles on the different video lists. }
      \vspace{-3mm}
  \begin{tabular}{l|c|ccc|ccc|ccc}
  \toprule
 \multicolumn{2}{c}{ } &\multicolumn{3}{|c}{\KidsList } & \multicolumn{3}{|c}{\SeedList } & \multicolumn{3}{|c}{\ControlList} \\
%\hline
\textbf{Profile}& \textbf{No. ads} & \textbf{Other} & \textbf{Placement}&  \textbf{Non-pers.}& \textbf{Other}& \textbf{Placement }& \textbf{Non-pers.} &  \textbf{Other} & \textbf{Placement}& \textbf{Non-pers.}  \\ \midrule
    P1 & 239	& 36 &\textbf{0}  & 129 & 26 & \textbf{0} & 	0 &  48& \textbf{0}  & 0 \\
    P2 & 205 &	10& \textbf{0}&		52& 68&	\textbf{0}&		0&	75&\textbf{0}&		0 \\
    P3 & 599 &	75& \textbf{27}&		274&	 83&\textbf{39}&		0&	55& \textbf{46}&		0 \\
    P4&	535 &	 114& \textbf{25}&		251&	46& \textbf{17}&		0&	55& \textbf{27}&		0 \\
    P5&	934 &	 144& \textbf{49}&		443&	112& \textbf{40}& 0&	122& \textbf{23}& 0 \\
    P6&	809 &	 \textbf{0}& \textbf{0}&		\textbf{390}&	\textbf{0}& \textbf{0}&		\textbf{264}&	110& \textbf{45}&		0 \\
 \arrayrulecolor{gray}\specialrule{0.3pt}{1pt}{1pt}  \arrayrulecolor{black}
    Total 	&3,321 &379	&\textbf{101} &1,539 &335	&\textbf{96}&264	&465&\textbf{141}		&0 \\
   \bottomrule
  \end{tabular}

  \label{tab:all results}
  }
\end{table*} 

\subsection{Analysis of ads}
\label{sec:res}

This section analyzes the 3,221 ads we collected with the six profiles when watching the 620 videos in YouTube.

\paragraph{Ad frequency}
Table~\ref{tab:ads} shows the number of video and float ads we collected with the different profiles.   We can see that P1 and P2, which do not have associated Google profiles, have not received any float ads and have the lowest number of video ads. For the other profiles, we can notice that they receive approximately the same number of ads. Still, the highest number of ads are for browser profiles associated with Google profiles with browsing history (P5 and P6). 

%\begin{table}
%  \caption{Number of ads collected from the six profiles. }
%  \label{tab:ads}
%\footnotesize{
%\begin{tabular}{ccc}
%\hline
%\textbf{Profile}  & \textbf{Number of Video ads} & \textbf{Number of Float ads}  \\ \hline
%P1 & 239 & 0 \\
%P2 & 205 & 0\\
%P3 & 437 & 162\\
%P4 & 396 & 139\\
%P5 & 541 & 393 \\
%P6 & 441 & 368  \\
%ALL & 2259 & 1062 \\
%\hline
%\end{tabular}
%}
%\end{table}

\paragraph{Overlap}
Table~\ref{tab:overlap} reports the overlap in video ads between the six browser profiles. Recall that all six browser profiles are watching the same videos at relatively the same time.   To compute the overlap, we computed the unique set of video ads for each profile and computed the intersection between the different profiles. We can see that the overlap between P1 and P2 is 19\% (the two browser profiles are not associated with Google profiles). In contrast, the intersection of either P1 or P2 with the rest of the profiles is much lower. The intersection between the four profiles associated with a Google profile, with or without browsing history, is between 15\% to 23\%. This suggests that the set of ads received by browser profiles not associated with a Google profile differs from browser profiles with an associated Google profile.

%
%P12 142 ? 119 ? 23 ? 19
%P13 313 ? 335 ? 23  ?6.8
%P14 313 ? 289 ? 24 ? 8.3 
%P15 326  ? 300 ? 26 ?8.6
%P16 265 ? 232 ? 33 ? 14
%P23 320 --  312 ? 8  -- 2.5 
%P24 275 ? 267 ? 8 --2.9
%P25 288 ? 280 ? 8 -- 2.8 
%P26 227 ? 216 ? 8  -- 3.7 
%P34 491 ? 411 ? 80 ? 19
%P35 504 ? 418 ? 86 ? 20.5
%P36 443 ? 376 ? 63 ? 16.7 
%P45 459 ? 395 ? 64 ?16.2
%P46  398 ? 348 ? 50 ? 14.3 
%P56 411 ? 334 ? 77 ? 23

%\begin{table*}[t]
%  \centering 
%    \footnotesize{
%      \caption{Overlap between video ads across different profiles.}
%  \begin{tabular}{c|c|c|c|c|c|c|c|c|c|c|c|c|c|c|c}
%\hline
%\textbf{Profiles}&P1--P2 & P1--P3 & P1--P4 & P1--P5 & P1--P6 & P2--P3 & P2--P4 & P2--P5 & P2--P6 & P3--P4 & P3--P5 & P3--P6 & P4--P5 & P4--P6 & P5--P6 \\ 
%\hline
%\textbf{Overlap}&\textbf{19\%}&6.8\%& 8.3\% &8.6\%& 14\%& 2.5\% &2.9\%&2.8\% &  3.7\% &\textbf{19\%}& \textbf{20.5\%}& \textbf{16.7\%} &\textbf{16.2\%}&\textbf{14.3\%} & \textbf{23\%} \\
%\hline
%   
%  \end{tabular}
%
%  \label{tab:all results}
%  }
%\end{table*} 

\paragraph{Placement-based ads}
The central question of this section is whether advertisers employ placement-based targeting on children-focused videos in the real-world. Table~\ref{tab:all results} presents the ad explanations corresponding to the ads received by the six browser profiles for videos on the \KidsList, the \SeedList, and the \ControlList. 
We consider an ad to be \textit{placement-based} if the ad explanation is ``\textit{The video you are watching}'' (i.e., EID~1). We consider an ad to be \textit{non-personalized} if the ad explanation is ``\textit{Ad personalizing is disabled for this account or content. Therefore, this ad is not personalized based on your data. Its distribution depends on other factors (such as the time or your geographical position)}'' (i.e., EID~27). And \textit{other} if the ad explanations are one of EID~2~to~26. 

The table shows that ads with a placement-based ad explanation account for 5\% of ads collected on \KidsList; 14\% of the ads collected on \SeedList; and 23\% of the ads collected on \ControlList.  
%In addition, the table shows that ads with a \textit{other} ad explanation account for 21\% if ads collected on \KidsList; 35\% on ads collected on \SeedList; and 77\% on ads collected on \ControlList. 
Hence, 7\% of ads across both our children-focused lists (\KidsList and \SeedList) have placement-based ad explanations, suggesting that \textbf{real-world advertisers use placement-based targeting on children-focused videos}. 
Given the limitations of ad explanations, the biased set of videos we watch, and our made-up browser profiles, this number should only be seen as evidence that such targeting occurs in practice without being representative.

%In addition, some of the ads shown on children-focused videos are not clearly marked as not being personalized. 

\paragraph{Non-personalized ads}
Table~\ref{tab:all results} shows that ads that 76\% of ads in the \KidsList, 38\% of ads in \SeedList and 0\% of ads in \SeedList have a non-personalized ad explanation. At a first sight this makes sense, as YouTube might impose stricter restrictions on videos and channels it had whitelisted to be part of YouTube Kids than other children-focused videos and adult-focused videos. 

However, what triggered our attention is that while the majority of ads on \KidsList have a non-personalized ad explanation, there are some that have a placement-based or other ad explanation. Hence, we manually checked whether there is any difference between videos with ads whit non-personalized explanations and videos with ads with other or placement-based explanations. We observed two things: (1) Not all videos in \KidsList had a redirection on YouTube to YouTube Kids as presented in Figure~\ref{fig:youtube-kids-redirection}). To clarify, while all the videos in \KidsList are accessible on YouTube Kids, not all have the YouTube Kids redirection on YouTube. (2) We observed that all the ads on videos with a YouTube Kids redirection have a non-personalized ad explanation. At the same time, none of the ads in \KidsList with no YouTube Kids redirection have a non-personalized ad explanation. 
This was surprising, as our experiments in Section~\ref{sec:ad_targeting} showed that we could target videos that appear on YouTube Kids, including the videos YouTube Kids redirection. Hence, we decided to test the logic behind the non-personalized ad explanation.

\paragraph{On the correctness of the non-personalized ad explanation}
We hypothesized that Google provides a non-personalized ad explanation for videos with the YouTube Kids redirection, irrespective of the actual targeting parameters specified by the advertiser and used to deliver the ad on the video.  
%To test this, we created an ad experiment to test whether the ad explanation provided by Google on videos with a YouTube Kids redirection correctly reflects the targeting parameters chosen by the advertiser.
To test this, we created an ad campaign, and we set as placement nine videos from \ControlList, five videos from \SeedList without a YouTube Kids redirection, and 15 videos from \KidsList with a YouTube Kids redirection. Statistics provided by Google showed that the ad had 466 impressions on \ControlList, 41 on \SeedList, and 372 on \KidsList. 

During the experiment, we used one single profile and watched de various videos used in the placement to collect our ad and the corresponding ad explanation provided by Google. Table~\ref{tab:ad compaign results} presents the ad explanations we collected for our ad. %, for the same ad campaign, with the same profile, across the different types of videos.  
We can see that the ad explanation we received on both the \ControlList videos and \SeedList videos is consistent with the configuration of our ad campaign (i.e., placement-based). However, the explanation we got on \KidsList indicates that this ad is not personalized. 
Hence, for the same ad campaign and browser profile, \textbf{Google changed its explanation depending on whether the video has a YouTube Kids redirection or not}. This situation is misleading and alarming; not only can children be targeted with ads through placement-based advertising, but the ad explanations provided on many children-focused videos do not reflect the use of this targeting technique. Hence, \textit{placement-based targeting might be even more frequent than we estimated on children-focused videos, as Google does not provide the precise parameters advertisers use to target their ads on many of the videos we tested.} 

Finally, the results for P6 are also strange because it has non-personalized ad explanations for all the ads in the children-focused videos we tested, while other profiles, such as P3, P4, and P5, do not. According to Table~\ref{tab:overlap} there is a 15\% to 20\% overlap between the ads received by P6 and ads received by P3, P4, or P5. We checked the common ads, and we observed they have different ad explanations when shown to P6 and when shown to the other profiles. We believe that Google detected activity consistent with a child watching videos and changed the ad explanation policy for P6. 
We believe that the current ad explanations provided by Google on children-focused videos are misleading and not ``meaningful''  as demanded in Article 26 of the DSA. 

\begin{table}
  \caption{Ad explanations provided by YouTube for our ad campaign by category of video.}
  \vspace{-3mm}
  \label{tab:ad compaign results}
  \footnotesize{
\begin{tabular}{p{7.5cm}}
\toprule
 \textbf{Ad explanation} \\ \midrule
\ControlList 

 ``\textbf{The video you are watching.}
  Your activity while you were logged in to Google.
   Google's estimate of your current approximate location.
   Google's estimate (based on your past activity) of general geographies you may be interested in."
    \\
 \arrayrulecolor{gray}\specialrule{0.3pt}{1pt}{1pt}  \arrayrulecolor{black}
\SeedList 

     ``\textbf{The video you are watching.}
     Your activity while you were logged in to Google.
     Google's estimate of your current approximate location.
     Google's estimate (based on your past activity) of general geographies you may be interested in."
  \\
 \arrayrulecolor{gray}\specialrule{0.3pt}{1pt}{1pt}  \arrayrulecolor{black}
\KidsList 

``\textbf{Ad personalizing is disabled for this account or content. Therefore, this ad is not personalized based on your data. Its distribution depends on other factors (such as the time or your geographical position).}'' \\
 \bottomrule
\end{tabular}
}
\end{table}

%!TEX root = main.tex
\section{Limitations}

\noindent \textbf{Ad explanations correctness} One important limitation of our study is that we rely on ad explanations provided by Google to detect the corresponding targeting parameters used by advertisers. As we showed these, explanations are not always an accurate reflection of the targeting parameters. 
In particular, because we could not fully prove specificity, the fraction of ads targeted using placement-based targeting might be lower if the ad explanation ``The video you are watching?'' is also given for other types of targeting. 
While we tested specificity with respect to four types of targeting (theme-based, behavioral-based, demographic-based, and general-based), proving complete specificity is hard  as even if we could manage to test all the different combinations of targeting parameters offered by ad platforms (which is financially prohibitive in our case), there could still be a theoretical chance, there are parameters outside our control that could trigger this ad explanation. 
The other problem comes from the fact that Google did not provide the precise targeting parameters on the videos we tested that had a YouTube Kids redirection. Hence, placement-based advertising might even be higher than reported. 
Nevertheless, we think that the paper's more long-lasting and meaningful contribution is pinpointing that placement-based targeting can be used to target children and show that the use of placement-based advertising on children-focused videos occurs in practice, irrespective of how prevalent this is.

%While we tested the correctness of some of these ad explanations (in particular for placement-based advertising), we could not test all possible combinations of targeting parameters and check the corresponding explanations.  
%This means that the fraction of ads targeted using placement-based targeting might be lower if the ad explanation ``The video you are watching?. is also given for other types of targeting. This would be hard to prove as  proving specificity\footnote{Specificity means that if we receive the ad explanation ?The video you are watching?, it always corresponds to placement-based targeting and not to other types of targeting.} is hard

%Hence, we tested the specificity with respect to four other types of targeting (theme-based, behavioral-based, demographic-based, and general-based), which is a subset of all possible targeting strategies, but covers some of the most used targeting techniques. For none of these four categories of targeting, we collected a ``The video you are watching'' explanation. This shows specificity with respect to these four categories of targeting. 

\vspace{1mm}
\noindent \textbf{Ad delivery statistics}  We also rely on statistics provided by Google to check if our ads were delivered on the videos we asked as placement and if our ads were delivered to audiences with the characteristics we asked for. We think Google does not have any incentives to provide inaccurate statistics as this data is key to businesses deploying ad campaigns on their platform. 

\vspace{1mm}
\noindent \textbf{Small dataset and representativeness} The videos in our children-focused lists are biased toward videos coming from channels recommended by Google, channels that are popular, and videos that are recent. The evaluation is also done over a small dataset that uses only six non-representative browser profiles to watch content.   Hence, our results on the prevalence of placement-based advertising are not representative of the fraction of placement-based ads children receive when watching videos on YouTube and should only be seen as evidence that this occurs in practice. 

\vspace{1mm}
\noindent \textbf{No real data} In this study, we did not use data collected from actual children. Such data collection would be much harder to get, as we would need to convince parents to install our tool. In addition, the data collection would pose many more ethical and legal issues. We leave the collection and analysis of such data for future works. 

\vspace{1mm}
\noindent \textbf{Targeting mechanisms} There might be other ways, besides placement-based targeting, to reach children on YouTube, such as keyword-based or theme-based targeting, that we did not investigate. We, again, leave this for future work.

%!TEX root = main.tex

\section{Related Works}
\label{sec:rw}

There is a large body of work on advertising and children. Many of the seminal studies have been done by child developmental psychologists and have focused on 
ads on television. Next, we present some works that look at a child's understanding, perception, and effect of ads.

In 1982, in the context of television, Stephen, Levin, and Petrella found that children as young as three years old could distinguish between TV programs and advertisements. Still, they had no understanding or awareness of advertisers' motivations and sales intentions~\cite{pretvad}. 
In 1992, Bijmolt et al.~\cite{Bijmolt98} further investigated the question and analyzed whether gender, age, and parental influence have an influence on whether children can distinguish between advertising content and other content, as well as understand its commercial intent. For this, they conducted verbal and non-verbal measurements on children between five and eight years old. This paper shows that, according to non-verbal measures, most children can distinguish commercials from programs and have some insight into advertising intent. Whereas verbal measures are not as conclusive, it shows that the percentage of children who show understanding of TV advertising is substantially lower. For the effect of age on understanding the ad, the study shows that the older the child is, the better they understand the ads. However, the effect of gender and parent-child interaction is relatively small.
Other works such as~\cite{ijrsi18} have shown that children tend to misunderstand the message of ads and focus more on the negative side of the message than the positive side. However, this work has also emphasized the potential positive effect of ads as children can be better informed on many subjects. For example, an advertisement for toothpaste can make them aware of problems with their teeth if one does not brush their teeth regularly.
Related to this, a children's psychology study claimed that young children do not realize that a message (of an ad) can only present positive information while retaining negative information to manipulate the mental state of people~\cite{AloiseYoung}. The study analyzed the development of self-promotion in 6- to 10-year-old children. Children were asked to submit a description of themselves to be chosen for a fictional team. Younger children included both negative and positive information. In comparison, older children were better at self-promotion and included only positive descriptions. Hence, older children can understand that ads only present positive information, while younger children do not. 
A large body of research and reports have specifically focused on the negative effects of food advertising on children~\cite{Story04,FOLKVORD201626,doi:10.2105/AJPH.2009.179267,https://doi.org/10.1111/pedi.12278,FOLTA2006244,kelly_smith_king_flood_bauman_2007,10.1542/peds.2016-1758V,10.1542/peds.2020-1681}.
According to all these studies, it is clear that we need to take measures to protect children from possible manipulation and harm from advertisers. 

A survey by Raju et al.~\cite{RajuLonial90} said that targeting children has been part of the corporate business since the appearance of ads.
Among the reasons that push a company to target a child: is the size of the business; there were 14 million children/teenagers between 0 and 17 years old in France in 2018, according to INSEE. The second reason would be the purchasing power of children who receive more and more money nowadays; their consumption is mainly focused on snacks and sweets, followed by toys and games. One of the other most important reasons remains the power of persuasion that a child can have on their parents; a child actively contributes to the purchasing choices in the family, from breakfast cereals to the color of the car to buy. In addition, the children of today are the adults of tomorrow and those who will be able to spend their own money. 
They also discussed that the content of advertisements could influence a child's consumption, especially when it is about food. Among other points, they concluded that advertisements targeting adults could influence children's vision of the world of adults, their preferences, and choices when they grow up.

A more recent body of work has studied online advertising and children~\cite{Austin1999TargetingCO,doi:10.1080/02650487.2016.1196904,doi:10.2501/IJA-33-3-437-473,doi:10.1080/17482791003629610,CAI20131510,Moore2004-MOOCAT}. A recent (2020) report from Common Sense~\cite{commonsense1} analyzed what videos children are watching on Youtube and what is being advertised to children. To gather children viewing data, they asked parents to participate in a study and provide the URLs
their children viewed by copying and pasting a list from the browsing history of YouTube~\cite{commonsense0}. 
They coded more than 1600 videos afterward according to the topic and analyzed the ads on these videos. They found that about 20\% of the videos contained age-inappropriate ads. 
Other works have looked at the frequency and duration of advertising on popular children-focused channels on YouTube~\cite{10.1001/jamanetworkopen.2021.9890}. Jenny Radesky encourages pediatricians to discuss with parents and children the dangers of advertising and explain children's cognitive capabilities when faced with ads~\cite{Radesky20}. 
Finally, three other works have looked at mobile applications used by children: one study provides a content analysis of advertising~\cite{Meyer2019AdvertisingIY}; a second study looked at data collection practices~\cite{Zhao20}, and a final study looked at the compliance of mobile applications with COPPA~\cite{Reyes18}. 
While many studies are emerging in recent years around online advertising~\cite{Silva20a,andreou2019measuring,andreou2018investigating,ribeiro2019@fat,edelson2020security,ali2019discrimination,ali2019ad}, to our knowledge, no study has investigated how advertisers can target children with ads on video streaming platforms and analyze the available targeting techniques from the perspective of current and future laws.

%!TEX root = main.tex
\section{Conclusion}

This paper investigates the mechanisms through which advertisers can target their ads to children on YouTube and whether real-world advertisers are actually employing such mechanisms. Our experimental methodology showed that advertisers could target children through placement-based targeting. This targeting technique allows advertisers to specify the videos on which they want their ads to be shown. Our measurement methodology provides evidence that supports that real-world advertisers employ these techniques to target children with ads. 
We also discovered that placement-based advertising can be combined with targeting based on user profiling.  

We performed a legal analysis of the different texts regulating children's advertising in the E.U. and the U.S. According to COPPA and DSA, placement-based advertising on children-focused videos is not forbidden; however, targeting children based on user profiling is restricted (in the case of COPPA) and forbidden (in the case of DSA). The DSA, which was published in October 2022, is supposed to come into force gradually, and platforms have until 2024 to comply. % We hope our study raises awareness of the possible targeting mechanisms that advertisers can exploit to target children and allows lawmakers to refine the legislation. \update{We hope platforms} 
As many countries are updating their legislation to protect children from online harm, we hope our study raises awareness of the possible targeting mechanisms that advertisers can exploit to target children and allows lawmakers to refine the legislation. %In particular, we want to bring to the attention of regulators that advertisers can reach children through placement-based advertising which is a pra; and whether such practice should be allowed given the potentially harmful effects it could have on children. 

%In addition, current online advertising legislation is missing clear guidelines on the content that can be targeted at children (as in the case of television advertising). This is especially important since current laws permit targeting children to placement-based targeting. 

Given the fact that targeting children is permitted through contextual advertising and that ad platforms allow advertisers to place an ad under a single piece of content specifically targeted at children (e.g., a particular cartoon or a particular video clip dedicated to children), this can make verification of the content of the advertisement very difficult in practice, as it will only be seen by children who have seen that content, with very little possibility of adequate verification of the content of the ad by adults. Therefore, we strongly believe that we need transparency mechanisms that allow researchers and the civil society to scrutinize these ads. For example, ad platforms could provide ad libraries for ads shown on children-focused videos as they do for political ads.

%More precisely, our experiments show that there is a need to better distinguish between cases where targeting is carried out solely on the basis of context or where targeting is carried out by mixing contextual placement and profiling. This distinction is necessary to effectively implement the European (DSA) and American (COPPA) texts protecting children against advertising based on profiling.

%To our surprise, we discovered that Google provides ad explanations that do not reflect the precise targeting parameters chosen by advertisers for ads that appear on some children-focused videos (while it does for ads that appear on other videos). This weakens consumers' trust and understanding of the system and hinders external audit capabilities. 

Google has set up a streaming platform dedicated to children, YouTube Kids, with increased ad and content restrictions to protect children. This is a very positive initiative. However, surveys are showing that YouTube might still be more used by parents; hence, we encourage Google to adopt the same restrictions on the main platform YouTube for children-focused videos and have efficient algorithms to detect children-focused videos that are not labeled as such by content creators.

\section{Acknowledgements}
We thank the anonymous reviewers for their helpful comments.
This research was supported in part by the French National Research Agency (ANR) through the ANR-17-CE23-0014, ANR-21-CE23-0031-02, by the MIAI@Grenoble
Alpes ANR-19-P3IA-0003 and by the EU 101041223, and 101021377  grants.

\bibliographystyle{ACM-Reference-Format}
\bibliography{sample.bib}

%\bibliographystyle{plain}
%\bibliography{sample.bib}
%!TEX root = main.tex

\begin{appendix}
\section{Appendix}
\label{appendix}

\subsection{Screenshots}
\label{app:scr}

\begin{figure}[ht]
        \centering
        \includegraphics[scale=0.33]{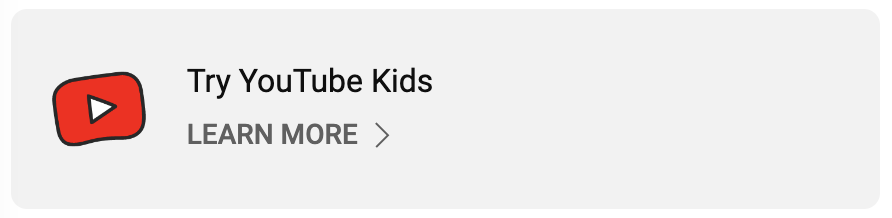}

        \caption{Redirection from YouTube to YouTube Kids.}
        \label{fig:youtube-kids-redirection}
        \vspace{-4mm}
    \end{figure}
    
    \begin{figure}[ht]
        \centering
        \includegraphics[scale=0.33]{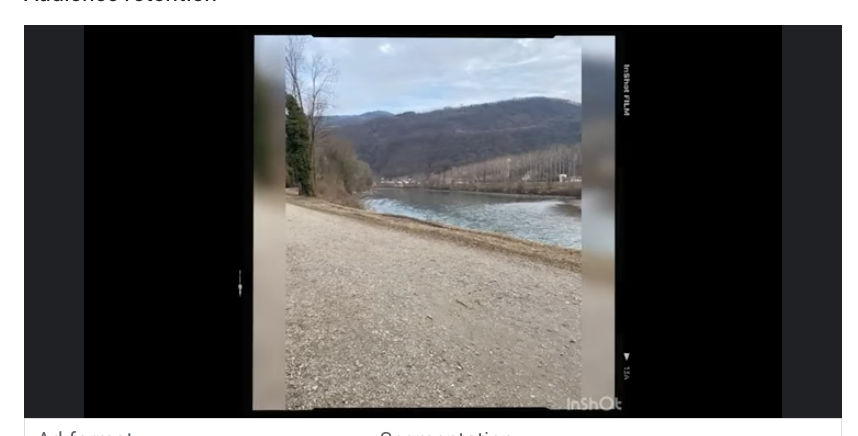}
        \hfill
        
                \includegraphics[scale=0.33]{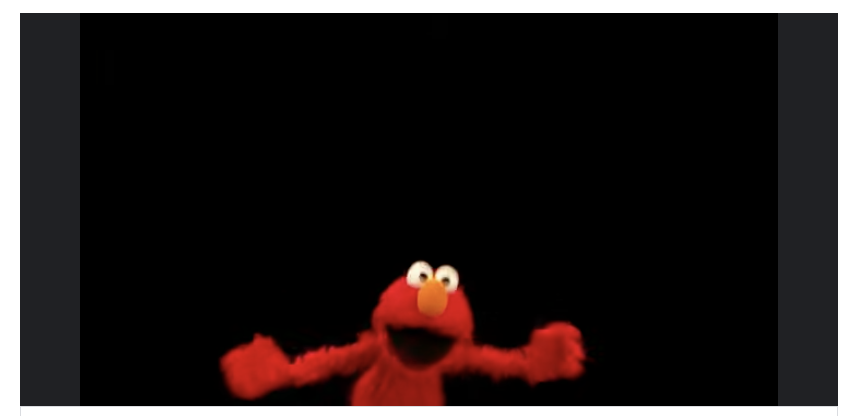}
        \caption{Screenshot of our two ads.}
        \label{fig:ad1}
        \vspace{-4mm}
    \end{figure}
    
%    \begin{figure}[ht]
%        \centering
%        \includegraphics[scale=0.33]{images/ad2.png}
%        \caption{Screenshot of our ad.}
%        \label{fig:ad2}
%        \vspace{-4mm}
%    \end{figure}

    \begin{figure}[t]
        \centering
        \includegraphics[scale=0.33]{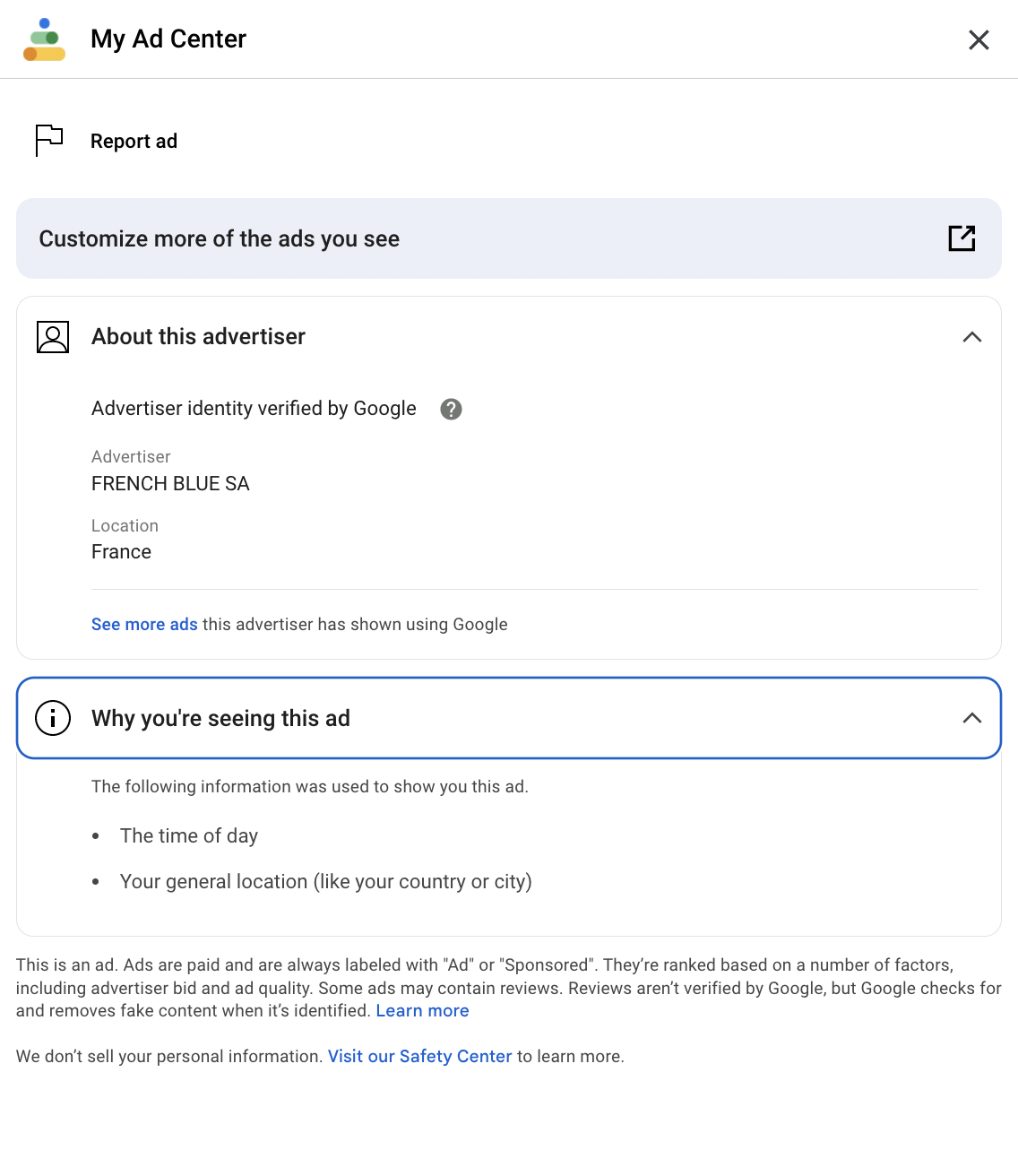}
        \vspace{-2mm}
        \caption{Example of ad explanation provided in ``Why you're seeing this ad''.}
        \label{fig:ad_explanaiton}
        \vspace{-6mm}
    \end{figure}
    
\subsection{Additional legal excerpts}
\label{app:law}
This section provides more details on the meaning and definition of precise words used in different legislations we discuss in Section~\ref{sec:bkg_law}.

\subsubsection{Additional provisions in the DSA}
The Article 28 of the DSA is without prejudice to Union law on protection of personal data. In particular, \underline{Article 8 of the GDPR} states, concerning conditions applicable to child's consent in relation to information society services, that, {\em ``where the criteria for making data processing lawful is consent (6.1.a of the GDPR) (...) the processing of the personal data of a child shall be lawful where the child is at least 16 years old. Where the child is below the age of 16 years, such processing shall be lawful only if and to the extent that consent is given or authorised by the holder of parental responsibility over the child.}
Member States may provide by law for a lower age for those purposes provided that such lower age is not below 13 years. %For example, in France, a child can give his consent, where he is at least 15 years old.
%The controller shall make reasonable efforts to verify in such cases that consent is given or authorised by the holder of parental responsibility over the child, taking into consideration available technology.
%Moreover, Article 8 of the GPDR shall not affect the general contract law of Member States such as the rules on the validity, formation or effect of a contract in relation to a child.

To ensure effective implementation of these due diligence obligations by platforms, the DSA defines a set of administrative sanctions  and monitoring measures  that will be implemented by the European Commission  in relation to VLOPS. In \underline{Article 54} supplements this by the possibility for a recipient of services who suffers damage as a result of the violation of a due diligence obligation to obtain compensation for his loss in: ``{\em Recipients of the service shall have the right to seek, in accordance with Union and national law, compensation from providers of intermediary services, in respect of any damage or loss suffered due to an infringement by those providers of their obligations under this Regulation}''.  This solution is a major novelty. It is important to specify that liability will be triggered by the occurrence of a systemic risk, for example concerning the mental health of a child.

\subsubsection{Additional provisions in the Directive (EU) 2018/1808 of 14 November 2018}
\underline{Article 28b (3)} states that ``{\em \textbf{appropriate measures} shall be determined in light of the nature of the content in question, the harm it may cause, the characteristics of the category of persons to be protected as well as the rights and legitimate interests at stake, including those of the video-sharing platform providers and the users having created or uploaded the content as well as the general public interest (...).
For the purposes of the protection of minors (...) the most harmful content shall be subject to the strictest \textbf{access control} measures.
Those measures shall consist of, as appropriate : (...) (f) establishing and operating \textbf{age verification} systems for users of video-sharing platforms with respect to content which may impair the physical, mental or moral development of minors; (...) (g) providing for \textbf{parental control} systems that are under the control of the end-user with respect to content which may impair the physical, mental or moral development of minors (...). }''

\subsubsection{Relevant nomenclature in the COPPA}

\textbf{Verifiable parental consent} means ``{\em any reasonable effort (taking into consideration available technology), including a request for authorization for future collection, use, and disclosure described in the notice, to ensure that a parent of a child receives notice of the operator's personal information collection, use, and disclosure practices, and authorizes the collection, use, and disclosure, as applicable, of personal information and the subsequent use of that information before that information is collected from that child.''}

\textbf{Personal information}  means ``\em{individually identifiable information about an individual collected online, including: 
(1) A first and last name; 
(2) A home or other physical address including street name and name of a city or town; 
(3) Online contact information as defined in this section; 
(4) A screen or user name where it functions in the same manner as online contact information, as defined in this section; 
(5) A telephone number; 
(6) A Social Security number; 
(7) A persistent identifier that can be used to recognize a user over time and across different Web sites or online services. Such persistent identifier includes, but is not limited to, a customer number held in a cookie, an Internet Protocol (IP) address, a processor or device serial number, or unique device identifier; 
(8) A photograph, video, or audio file where such file contains a child's image or voice; 
(9) Geolocation information sufficient to identify street name and name of a city or town; or 
(10) Information concerning the child or the parents of that child that the operator collects online from the child and combines with an identifier described in this definition.}''

\end{appendix}

\end{document}